\documentclass{article}[11pt]

\pdfoutput=1

\usepackage{a4wide}
\usepackage[top=3cm]{geometry}

\usepackage{authblk}

\usepackage{graphicx}
\usepackage{latexsym}

\usepackage{array}
\usepackage{url}
\usepackage{multirow}
\usepackage{multicol}
\usepackage{here}
\usepackage{comment}

\usepackage{amsthm}
\usepackage{amsmath,amssymb}

\newcommand{\Suffix}{\mathit{Suffix}}
\newcommand{\Substr}{\mathit{Substr}}

\newcommand{\STree}{\mathsf{STree}}
\newcommand{\DAWG}{\mathsf{DAWG}}
\newcommand{\CDAWG}{\mathsf{CDAWG}}

\newcommand{\SA}{\mathsf{SA}}
\newcommand{\microtree}{\mathsf{mt}}

\newcommand{\suflink}{\mathit{slink}}

\newcommand{\rev}[1]{\overline{#1}}

\def\W#1#2{{\cal W}_{#1}({#2})}

\newcommand{\Fpathstr}{\mathit{str}_{\mathsf{f}}}
\newcommand{\Bpathstr}{\mathit{str}_{\mathsf{b}}}

\newcommand{\FT}{\mathsf{T}_{\mathsf{f}}}
\newcommand{\BT}{\mathsf{T}_{\mathsf{b}}}
\newcommand{\FV}{\mathsf{V}_{\mathsf{f}}}
\newcommand{\BV}{\mathsf{V}_{\mathsf{b}}}
\newcommand{\FE}{\mathsf{E}_{\mathsf{f}}}
\newcommand{\BE}{\mathsf{E}_{\mathsf{b}}}

\newcommand{\anc}{\mathit{anc}}
\newcommand{\des}{\mathit{des}}

\newcommand{\Frmaximal}{\mathit{r\mathchar`-mxml}_{\mathsf{f}}}
\newcommand{\Flmaximal}{\mathit{l\mathchar`-mxml}_{\mathsf{f}}}
\newcommand{\Fmaximal}{\mathit{mxml}_{\mathsf{f}}}
\newcommand{\Blmaximal}{\mathit{l\mathchar`-mxml}_{\mathsf{b}}}
\newcommand{\Brmaximal}{\mathit{r\mathchar`-mxml}_{\mathsf{b}}}
\newcommand{\Bmaximal}{\mathit{mxml}_{\mathsf{b}}}

\newcommand{\occ}{\mathit{occ}}
\newcommand{\lca}{\mathit{lca}}

\newcommand{\FEeqc}[1]{[#1]_{E, \mathsf{f}}}

\newcommand{\BEeqc}[1]{[#1]_{E, \mathsf{b}}}

\newcommand{\NumNode}[1]{|#1|_{\#{\mathit{Node}}}}
\newcommand{\NumEdge}[1]{|#1|_{\#{\mathit{Edge}}}}

\newtheorem{theorem}{Theorem}
\newtheorem{lemma}{Lemma}
\newtheorem{corollary}{Corollary}
\newtheorem{fact}{Fact}

\theoremstyle{remark}
\newtheorem*{remark}{Remark}

\begin{document}

\title{Towards a complete perspective on labeled tree indexing: \\ new size bounds, efficient constructions, and beyond}

\author{Shunsuke Inenaga}

\affil{
  \normalsize{
  \textit{Department of Informatics, Kyushu University, Fukuoka, Japan}\\
  \textit{PRESTO, Japan Science and Technology Agency, Kawaguchi, Japan}\\
  \texttt{inenaga@inf.kyushu-u.ac.jp}
  }
}

\date{}

\maketitle

\begin{abstract}
  A labeled tree (or a trie) is a natural generalization of a string,
  which can also be seen as a compact representation of a set of strings.
  This paper considers the labeled tree indexing problem,
  and provides a number of new results on space bound analysis, and
  on algorithms for efficient construction and pattern matching queries.
  Kosaraju [FOCS 1989] was the first to consider
  the labeled tree indexing problem, and he proposed
  the suffix tree for a backward trie,
  where the strings in the trie are read in the leaf-to-root direction.
  In contrast to a backward trie, we call a usual trie as a forward trie.
  Despite a few follow-up works after Kosaraju's paper,
  indexing forward/backward tries is not well understood yet.
  In this paper, we show a full perspective on the sizes of indexing structures
  such as suffix trees, DAWGs, CDAWGs, suffix arrays, affix trees, affix arrays
  for forward and backward tries.
  Some of them take $O(n)$ space in the size $n$ of the input trie,
  while the others can occupy $O(n^2)$ space in the worst case.
  In particular, we show that the size of the DAWG for a forward trie
  with $n$ nodes
  is $\Omega(\sigma n)$, where $\sigma$ is the number of distinct characters in the trie.
  This becomes $\Omega(n^2)$ for an alphabet of size $\sigma = \Theta(n)$.
  Still, we show that there is a compact $O(n)$-space implicit representation of the DAWG
  for a forward trie, whose space requirement is independent of the alphabet size.
  This compact representation allows for simulating
  each DAWG edge traversal in $O(\log \sigma)$ time,
  and can be constructed in $O(n)$ time and space
  over any integer alphabet of size $O(n)$.
  In addition, this readily extends to the \emph{first} indexing structure
  that permits \emph{bidirectional pattern searches} over a trie
  within linear space in the input trie size.
  We also discuss the size of the DAWG built on a \emph{labeled DAG}
  or on an \emph{acyclic DFA}, and present a quadratic lower bound for its size. \\

\end{abstract}

\section{Introduction}

\emph{Strings} are an abstract data type for any data in which
the order of the items matters.
Strings cover a wide range of sequential data,
e.g., natural language text, biological sequences,
temporal data, time series, event sequences, and server logs.
Due to recent developments of sensor networks,
M2M communications, and high-throughput sequencing technologies, 
string data have been increasing more rapidly than ever before.

String indexing is a fundamental problem in theoretical computer science,
where the task is to preprocess a text string so that subsequent pattern matching
queries on the text can be answered quickly.
This is the basis of today's large-scale information retrieval systems and
databases.
The first of such string indexing structures was the suffix tree,
which was introduced by Weiner in 1973~\cite{Weiner}.
Suffix trees also have numerous other applications including
string comparisons~\cite{Weiner},
data compression~\cite{ApostolicoL00}, 
data mining~\cite{Muthukrishnan02},
and bioinformatics~\cite{gusfield97:_algor_strin_trees_sequen,MBCT2015}.

A \emph{labeled tree} is a static rooted tree
where each edge is labeled by a single character.
A \emph{trie} is a kind of labeled tree such that
the out-going edges of each node are labeled by mutually distinct characters.
A trie can be naturally seen as an \emph{acyclic deterministic finite-state automaton}
(\emph{acyclic DFA}) in a tree shape, which accepts a finite set of strings.
In this sense, a trie can be seen as a compact representation of a set of strings.
In another view, a trie is a generalization of a string
that is a labeled single-path tree.

This paper considers the \emph{labeled tree indexing problem},
where the task is to build a data structure that supports
sub-path queries to report all sub-paths in the trie
that match a given string pattern.
Such sub-path queries on labeled trees are primitive in data base searches
where the XML has been used as a \emph{de facto} format of data storage.
It is well known that each XML document forms a tree\footnote{Usually, a string label is associated to a \emph{node} in an XML tree structure. However, for such a tree, there is an obvious corresponding trie where each \emph{edge} is labeled by a single character.}.
Other important uses of labeled trees includes
SQL queries~\cite{Hammer004},
dictionaries~\cite{AC75,WittenMB99},
and data compression~\cite{LZ78,LZW}, to mention just a few.

A \emph{backward} trie is an edge-reversed trie,
where we read the path strings in the leaf-to-root direction.
Kosaraju~\cite{Kosaraju89a} was the first to consider
the trie indexing problem, and he proposed the suffix tree
of a backward trie that takes $O(n)$ space\footnote{We evaluate the space usage of algorithms and data structures by the number of machine words (not bits) unless otherwise stated},
where $n$ is the number of nodes in the backward trie.
Kosaraju also claimed an $O(n \log n)$-time construction.
Breslauer~\cite{breslauer_suffix_tree_tree_1998} showed
how to build the suffix tree of a backward trie in $O(\sigma n)$
time and space, where $\sigma$ is the alphabet size.
Shibuya~\cite{Shibuya03} presented an $O(n)$-time and space
construction for the suffix tree of a backward trie over an integer alphabet
of size $\sigma = O(n)$.
This line of research has been followed 
by the invention of 
XBWTs~\cite{FerraginaLMM09}, suffix arrays~\cite{FerraginaLMM09},
enhanced suffix arrays~\cite{KimuraK12},
and position heaps~\cite{NakashimaIIBT12} for backward tries.

\subsection{Suffix Trees, DAWGs, and CDAWGs for Forward/\\Backward Tries}

In addition to the suffix trees,
we also consider the \emph{directed acyclic word graphs} (\emph{DAWGs})
and the \emph{compact DAWGs} (\emph{CDAWGs}),
which were introduced by Blumer et al. in 1985~\cite{blumer85:_small_autom_recog_subwor_text},
 and in 1987~\cite{Blumer87}, respectively.
Similar to suffix trees,
DAWGs and CDAWGs support linear-time pattern matching queries.
For any string of length $m$,
the suffix tree, the DAWG, and the CDAWG contain linear $O(m)$ nodes and $O(m)$ edges~\cite{Weiner,blumer85:_small_autom_recog_subwor_text,Blumer87}.
These bounds have been generalized to a set of strings such that,
for any set of strings of total length $M$,
the (generalized) suffix tree, the DAWG, and the CDAWG
contain linear $O(M)$ nodes and $O(M)$ edges~\cite{Blumer87,GusfieldS04}.
All these bounds are independent of the alphabet size $\sigma$.

The DAWG of a string $w$ is the smallest DFA that accepts
the suffixes of $w$~\cite{blumer85:_small_autom_recog_subwor_text}.
DAWGs for strings have important applications including
pattern matching with don't cares~\cite{KucherovR97},
online Lempel-Ziv factorization in compact space~\cite{YamamotoIBIT14},
finding minimal absent words in optimal time~\cite{FujishigeTIBT16},
and dynamic multiple pattern matching~\cite{HendrianIYS19}.
The CDAWG of a string $w$ can be regarded as \emph{grammar compression} for $w$
and can be stored in $O(e)$ space, where $e$ denotes the number of
right-extensions of maximal repeats in $w$
which can be much smaller than the string length in highly repetitive strings~\cite{BelazzouguiC17}.
There is also an space-efficient suffix tree representation based
on the CDAWG~\cite{BelazzouguiC17}.
Hence, understanding DAWGs and CDAWGs for labeled trees is very important
and will likely lead to further advances in efficient processing of labeled trees.

To this end, this paper initiates size analysis on these indexing structures
for a forward (ordinary) trie and a backward trie.
We show that, quite interestingly, some of the aforementioned size bounds
\emph{do not} generalize to the case of tries.
We present \emph{tight} lower and upper bounds on the sizes of all these indexing structures,
as summarized in Table~\ref{tab:summary_size}.
Our size analysis is based on combinatorial properties that reside in
these indexing structures, such as the \emph{duality} of suffix trees and DAWGs,
and \emph{maximal repeats} on input tries,
and is not straightforward from the known bounds for strings.

\begin{table}[ht]
  \begin{center}
  \begin{tabular}{|c||c|c|c|c|} \hline
    \multicolumn{1}{|c||}{} & \multicolumn{2}{|c|}{forward trie} & \multicolumn{2}{|c|}{backward trie} \\ \hline
    index structure & \# of nodes & \# of edges & \# of nodes & \# of edges \\ \hline \hline
    suffix tree & \boldmath $O(n^2)$ & \boldmath $O(n^2)$ & $O(n)$ & $O(n)$ \\ \hline
    DAWG & $O(n)$ & \boldmath $O(\sigma n)$ & \boldmath $O(n^2)$ & \boldmath $O(n^2)$ \\ \hline
    CDAWG & \boldmath $O(n)$ & \boldmath $O(\sigma n)$ & \boldmath $O(n)$ & \boldmath $O(n)$ \\ \hline
    suffix array & \boldmath $O(n^2)$ & \boldmath $O(n^2)$ & $O(n)$ & $O(n)$ \\ \hline
  \end{tabular}
  \end{center}
  \vspace*{2mm}
  \caption{Summary of the numbers of nodes and edges of the suffix tree, DAWG, and CDAWG for a forward/backward trie with $n$ nodes over an alphabet of size $\sigma$. The new bounds obtained in this paper are highlighted in bold. All the bounds here are valid with any alphabet size $\sigma$ ranging from $\Theta(1)$ to $\Theta(n)$. Also, all these upper bounds are tight in the sense that there are matching lower bounds (see Section~\ref{sec:new_bounds})}
  \label{tab:summary_size}
\end{table}

Let $n$ denote the number of nodes in a given forward trie
and in the corresponding backward trie.
Our new bounds are summarized as follows:
\begin{itemize}
\item We first present a (folklore) result such that the number of nodes and
  the number of edges of the suffix tree for a forward trie are both $O(n^2)$
  for any forward tries,
  and are $\Omega(n^2)$ for some tries.
  These bounds are independent of the alphabet size $\sigma$.

\item As direct consequences to the aforementioned results on suffix trees,
the sizes of the \emph{suffix arrays} for a forward trie
and a backward trie are $O(n^2)$ and $O(n)$, respectively.
These upper bounds are also tight.

 \item The number of nodes in the DAWG for a forward trie is known to be $O(n)$~\cite{MohriMW09},
however, it was left open how many edges the DAWG can contain.
It is trivially upper bounded by $O(\sigma n)$,
since any node can have at most $\sigma$ children in DAWGs.
We show that this upper bound is tight by presenting
a worst-case instance that gives $\Omega(\sigma n)$ edges in the DAWG for a forward trie.
Since this lower bound is valid for alphabet size $\sigma$ from $\Theta(1)$ to $\Theta(n)$,
we obtain an $\Omega(n^2)$ worst-case size bound for the DAWG of a forward trie.

 \item We show that the DAWG of a backward trie shares the same nodes with
    the suffix tree of the corresponding forward trie under reversal of substrings.
    This immediately leads to $O(n^2)$ and $\Omega(n^2)$ bounds
    for the numbers of nodes and edges of the DAWG for a backward trie,
    independently of the alphabet size.
  
  \item The CDAWG of a forward trie and the CDAWG of its corresponding backward trie
    also share the same nodes under reversal of substrings.
    This leads us to $O(n)$ bounds for the numbers of nodes
    in both of these CDAWGs.
    However, the number of edges can differ by at most a factor of $n$:
    The CDAWG of a forward trie contains $O(\sigma n)$ and $\Omega(\sigma n)$ edges
    in the worst case,
    but the CDAWG of a backward trie contains only $O(n)$ edges independently of the alphabet size.
    We remark that the $\Omega(\sigma n)$ lower bound for the CDAWG edges
    for the forward trie
    is valid for alphabet size $\sigma$ ranging from $\Theta(1)$ to $\Theta(n)$,
    and hence, it can contain $\Omega(n^2)$ edges in the worst case.
\end{itemize}

\subsection{Implicit $O(n)$-size Representation of the DAWG for Forward Trie}
Probably the most interesting result in our size bounds is
the $\Omega(n^2)$ lower bound for the size of the DAWG
for a forward trie with $n$ nodes over an alphabet of size $\Theta(n)$
(Theorem~\ref{theo:DAWG_FT_edge}):
Mohri et al.~\cite{MohriMW09} proposed an algorithm 
that constructs the DAWG for a forward trie with $n$ nodes
in time linear in the \emph{output} size.
Following our $\Omega(n^2)$-size lower bound, 
Mohri et al.'s construction must take at least $\Omega(n^2)$
time and space in the worst case.

Now, one may wonder whether or not it is possible to store
the DAWG for a forward trie in a compact manner, within linear space,
in the size of the input trie, in case of large alphabets.
Somewhat surprisingly, the answer to this challenging question is positive.
In this paper, we propose an \emph{implicit compact representation} of
the DAWG for a forward trie
that occupies only $O(n)$ space independently of the alphabet size,
and allows for simulating traversal of each DAWG edge in $O(\log \sigma)$ time.
We emphasize that this is the \emph{first} linear-space representation
of \emph{any} DFA that accepts all substrings of a given labeled tree,
after 35 years from the seminal paper by Blumer et al. in 1985 for the DAWG of a string~\cite{blumer85:_small_autom_recog_subwor_text}.
In addition, we present an algorithm that builds this implicit representation of the DAWG
for a forward trie in $O(n)$ time and space
for any integer alphabet of size $O(n)$.
Our data structure of an implicit representation of the DAWG
is carefully designed upon combinatorial properties of
\emph{Weiner links} that are defined on the
suffix tree of the corresponding backward trie.
Also, our algorithm does not require use of any additional complicated data structures.

\subsection{Relation to Indexing Automata and Labeled DAGs}
\label{sec:labaled_DAG}

Mohri et al.~\cite{MohriMW09} considered the problem of indexing
a given acyclic DFA (or equivalently a labeled DAG).
They proposed DAWGs (a.k.a. suffix automata) for acyclic DFAs,
and claimed that the number of \emph{nodes} in the DAWG for a given acyclic DFA with
$n$ states is $O(n)$.
However, they did not consider the number of \emph{edges} in the DAWG of an acyclic DFA.

A \emph{sequence binary decision diagram} (\emph{SDD})~\cite{LoekitoBP10}
is a family of Zero-Suppressed BDD (ZDD)~\cite{Minato93}
that represents a set of strings.
Roughly speaking, an SDD is a simple representation of a labeled DAG 
where the branches are implemented by a list that begins with
the leftmost child and continues to the right siblings.
SDDs are known to support a rich class of manipulations to the set of strings,
which can quickly be performed in practice.
Denzumi et al.~\cite{DenzumiTAM13} proposed an SDD-based indexing structure,
named \emph{PosFSDDdag}, for an input labeled DAG which is also
given in an SDD form.
PosFSDDdags can be regarded as a representation of DAWGs or suffix automata
for acyclic DFAs, with the leftmost-child-and-right-siblings representation of branches.
PosFSDDdags can be quite space-efficient in practice,
but theoretical size bounds of PosFSDDdag were not well understood in the literature~\cite{DenzumiTAM13}.

Notice that forward tries are the simplest kind of acyclic DFAs.
Thus, our analysis also resolves the aforementioned
open questions in terms of the worst case bounds.
Namely, our $\Omega(n^2)$ size bound for the DAWG of a forward trie
with $n$ nodes over an alphabet of size $\sigma = \Theta(n)$ immediately leads to
a worst-case quadratic size for DAWGs for input acyclic DFAs.
Figure~\ref{fig:DAWG_FT_edge_lowerbound} illustrates
a lower bound instance for DAWGs.
The number of edges of the broom-like input trie in Figure~\ref{fig:DAWG_FT_edge_lowerbound}
must be preserved in its SDD representation.
Moreover, even if we merge the sink nodes of the DAWG in Figure~\ref{fig:DAWG_FT_edge_lowerbound} into a single sink,
the number of edges in the DAWG still does not decrease.
Thus the worst-case size of PosFSDDdags is at least $\Omega(n^2)$
for some input SDD of size $n$.

\subsection{Bidirectional Pattern Search on Tries}
\label{sec:bidirectional_intro}

In the bidirectional pattern search problem,
the characters of a query pattern is given in a \emph{bidirectional online manner}, added either to the left end or to the right end of the current pattern.
This enables us to perform very flexible pattern searches so that given a current pattern,
one can look for its left or right contexts in a dynamic manner.
Bidirectional pattern searches in strings
have other various applications, including:
high-throughput short read alignments~\cite{LamLTWWY09},
discovery of secondary structures in RNA sequences~\cite{MauriP05,Strothmann07,SchnattingerOG12,GogKKMV14},
multiple pattern matching~\cite{GogKKMV14},
and approximate pattern matching~\cite{KucherovST16}.

Every existing bidirectional indexing structure for a string
consists of a pair of two indexing structures,
one built on a forward string (to be read from left to right) and the other on the corresponding backward string (to be read from right to left).
\emph{Affix trees}, proposed by Stoye in 2000~\cite{Stoye00},
are the first bidirectional indexing structure for strings,
followed by the \emph{affix arrays}~\cite{Strothmann07} counterparts.
\emph{Bidirectional BWTs}~\cite{SchnattingerOG12} are the most widely used bidirectional search structure nowadays, since they can be stored in compact or succinct space~\cite{BelazzouguiCKM13,BelazzouguiC19}.

Because a huge set of biological sequences can be compactly stored
as a tree in a practical database system~\cite{BieganskiRCR94},
bidirectional pattern search algorithms that work on tries are of high significance.
However, it seems that none of these existing bidirectional indexing structures
can readily be generalized to tries.
See also Table~\ref{tab:summary_size}.
Since affix trees (resp. affix arrays) contain at least the same number
of nodes (resp. entries) as the corresponding suffix trees (resp. suffix arrays),
the sizes of the affix tree/array built on
a pair of froward and backward tries turn out to be $O(n^2)$ and $\Omega(n^2)$ in the worst case.
It is not obvious if one can generalize BWT to a forward trie either,
because the corresponding suffix array contains $O(n^2)$ entries.

We focus our attention to suffix trees and DAWGs.
Using the duality of suffix trees and DAWGs~\cite{Blumer87},
one can indeed perform bidirectional pattern searches.
Therefore, in the case where $\sigma$ is a constant
(e.g., $\sigma = 4$ for DNA/RNA sequences),
the combination of the DAWG for a forward trie and
the suffix tree for a backward trie readily gives us
a $O(n)$-space bidirectional index for a trie.
However, when the underlying alphabet is very large,
which is common in temporal sequences (time series) and Japanese/Chinese texts, 
the $O(\sigma n)$-space requirement is prohibitive.
Still, our $O(n)$-space implicit representation of the DAWG for a forward trie
enables efficient bidirectional pattern searches on a trie for a large alphabet of size $\sigma = O(n)$.
To our knowledge, this is the \emph{first} linear-space bidirectional indexing
structure for labeled trees, after 20 years from the invention of
affix trees for strings in 2000~\cite{Stoye00}.

A portion of the results reported in this paper appeared in~\cite{Inenaga20}.

\section{Preliminaries}

Let $\Sigma$ be an ordered alphabet.
Any element of $\Sigma^*$ is called a \emph{string}.
For any string $S$, let $|S|$ denote its length.
Let $\varepsilon$ be the empty string, namely, $|\varepsilon| = 0$.
Let $\Sigma^+ = \Sigma^* \setminus \{\varepsilon\}$.
If $S = XYZ$, then $X$, $Y$, and $Z$ are called 
a \emph{prefix}, a \emph{substring}, and a \emph{suffix} of $S$, respectively.
For any $1 \leq i \leq j \leq |S|$,
let $S[i..j]$ denote the substring of $S$ that begins at position $i$
and ends at position $j$ in $S$.
For convenience, let $S[i..j] = \varepsilon$ if $i > j$.
For any $1 \leq i \leq |S|$, let $S[i]$ denote the $i$th character of $S$.
For any string $S$, let $\rev{S}$ denote the reversed string of $S$,
i.e., $\rev{S} = S[|S|] \cdots S[1]$.
Also, for any set $\mathbf{S}$ of strings,
let $\rev{\mathbf{S}}$ denote the set of the reversed strings of $\mathbf{S}$,
namely, $\rev{\mathbf{S}} = \{\rev{S} \mid S \in \mathbf{S}\}$.

\begin{figure}[htb]
        \centerline{
          \includegraphics[scale=0.6]{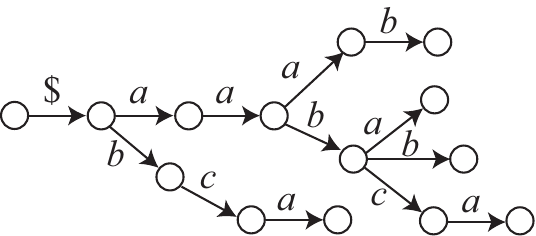}
          \hfil
          \includegraphics[scale=0.6]{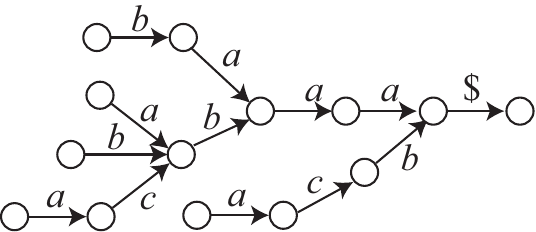}
        }
	\caption{A forward trie $\FT$ (left) and its corresponding backward trie $\BT$ (right).
          }
	\label{fig:FT_and_BT}
\end{figure}

A \emph{trie} $\mathsf{T}$ is a rooted tree $(\mathsf{V}, \mathsf{E})$ such that 
(1) each edge in $\mathsf{E}$ is labeled by a single character from $\Sigma$ and
(2) the character labels of the out-going edges of each node
begin with mutually distinct characters.
%
In this paper, a \emph{forward trie} refers to an (ordinary) trie
as defined above.
On the other hand, a \emph{backward trie} refers to an edge-reversed trie
where each path label is read in the leaf-to-root direction.
We will denote by $\FT = (\FV, \FE)$ a forward trie
and by $\BT  = (\BV, \BE)$
the backward trie that is obtained by reversing the edges of $\FT$.
We denote by a triple $(u, a, v)_{\mathsf{f}}$ an edge in a forward trie $\FT$,
where $u, v \in \mathsf{V}$ and $a \in \Sigma$.
Each reversed edge in $\BT$ is denoted by a triple $(v, a, u)_{\mathsf{b}}$.
Namely, there is a directed labeled edge $(u, a, v)_{\mathsf{f}} \in \FE$
iff there is a reversed directed labeled edge $(v, a, u)_{\mathsf{b}} \in \BE$.
See Figure~\ref{fig:FT_and_BT} for examples of $\FT$ and $\BT$.

For a node $u$ of a forward trie $\FT$,
let $\anc(u, j)$ denote the $j$th ancestor of $u$ in $\FT$ if it exists.
Alternatively, for a node $v$ of a backward $\BT$,
let $\des(v, j)$ denote the $j$th descendant of $v$ in $\BT$ if it exists.
We use a \emph{level ancestor} data structure~\cite{BenderF04}
on $\FT$ (resp. $\BT$) so that 
$\anc(u, j)$ (resp. $\des(v, j)$) can be found in $O(1)$ time
for any node and integer $j$, with linear space.
%

For nodes $u, v$ in a forward trie $\FT$ s.t. $u$ is an ancestor of $v$,
let $\Fpathstr(u, v)$ denote the string spelled out by
the path from $u$ to $v$ in $\FT$.
Let $r$ denote the root of $\FT$
and $\mathsf{L_f}$ the set of leaves in $\FT$.
The sets of substrings and suffixes of the forward trie $\FT$
are respectively defined by
\begin{eqnarray*}
  \Substr(\FT) & = & \{\Fpathstr(u, v) \mid u, v \in \FV, \mbox{$u$ is an ancestor of $v$}\}, \\
  \Suffix(\FT) & = & \{\Fpathstr(u, l) \mid u \in \FV, \mbox{$l \in \mathsf{L_f}$}\}.
\end{eqnarray*}

For nodes $v, u$ in a backward trie $\BT$ s.t. $v$ is a descendant of $u$,
let $\Bpathstr( v, u )$ denote the string spelled out
by the reversed path from $v$ to $u$ in $\BT$.
The sets of substrings and suffixes of the backward trie $\BT$
are respectively defined by
\begin{eqnarray*}
  \Substr(\BT) & = & \{\Bpathstr( v, u ) \mid v, u \in \BV, \  \mbox{$v$ is a descendant of $u$}\}, \\
  \Suffix(\BT) & = & \{\Bpathstr( v, r ) \mid v \in \BV, \ \mbox{$r$ is the root of $\BT$}\}.
\end{eqnarray*}

In what follows, let $n$ be the number of nodes in $\FT$ (or equivalently in $\BT$).
\begin{fact} \label{fact:FT_BT_subst_suffix}
(a) For any $\FT$ and $\BT$, $\Substr(\FT) = \rev{\Substr(\BT)}$.
(b) For any forward trie $\FT$, $|\Suffix(\FT)| = O(n^2)$.
    For some forward trie $\FT$, $|\Suffix(\FT)| = \Omega(n^2)$. 
(c) $|\Suffix(\BT)| \leq n-1$ for any backward trie $\BT$.
\end{fact}

\begin{figure}[htb]
  \centering
    \includegraphics[scale=0.6]{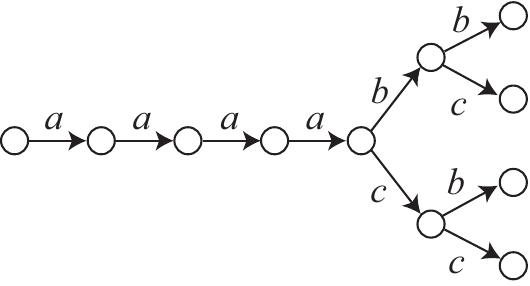}
	\caption{
          Forward trie $\FT$ containing distinct suffixes $a^i\{b, c\}^{\log_2(\frac{n+1}{3})}$ for all $i$~($0 \leq i \leq k = (n+1)/3$), which sums up to $k(k+1) = \Omega(n^2)$ distinct suffixes. In this example $k = 4$.
        }
  \label{fig:FT_worstcase}
\end{figure}

%
Fact~\ref{fact:FT_BT_subst_suffix}-(a), Fact~\ref{fact:FT_BT_subst_suffix}-(c)
and the upper bound of Fact~\ref{fact:FT_BT_subst_suffix}-(b)
should be clear from the definitions.
To see the lower bound of Fact~\ref{fact:FT_BT_subst_suffix}-(b) in detail,
consider a forward trie $\FT$ with root $r$ such that
there is a single path of length $k$ from $r$ to a node $v$,
and there is a complete binary tree rooted at $v$ with $k$ leaves
(see also Figure~\ref{fig:FT_worstcase}).
Then, for all nodes $u$ in the path from $r$ to $v$,
the total number of strings in the set
$\{\Fpathstr(u, l) \mid \mbox{$l \in \mathsf{L_f}$}\} \subset \Suffix(\FT)$
is at least $k(k+1)$,
since each $\Fpathstr(u, l)$ is distinct for each path $(u, l)$.
By setting $k \approx n/3$ so that the number $|\FV|$ of nodes in $\FT$ equals $n$,
we obtain Fact~\ref{fact:FT_BT_subst_suffix}-(b).
The lower bound is valid for alphabets of size $\sigma$ ranging from
$2$ to $\Theta(k) = \Theta(n)$.

\begin{remark}
  In case the input labeled tree is given as a compact tree
  (a.k.a. Patricia tree) where each edge is labeled by a non-empty string
  and every internal node is branching,
  then the input size $n$ should be translated to
  the total length of the string labels in the compact tree.
  Then all of the the following bounds also hold for input compact tries of size $n$.
\end{remark}

\section{Maximal Substrings in Forward/Backward Tries}

Blumer et al.~\cite{Blumer87} introduced the notions of
right-maximal, left-maximal, and maximal substrings in a set $\mathbf{S}$ of strings,
and presented clean relationships between the right-maximal/left-maximal/maximal substrings
and the suffix trees/DAWGs/\\CDAWGs for $\mathbf{S}$.
Here we give natural extensions of these notions to substrings in our 
forward and backward tries $\FT$ and $\BT$,
which will be the basis of our indexing structures for $\FT$ and $\BT$.

\subsection{Maximal Substrings on Forward Tries}
For any substring $X$ in a forward trie $\FT$,
$X$ is said to be \emph{right-maximal} on $\FT$ if
(i) there are at least two distinct characters $a, b \in \Sigma$
    such that $Xa, Xb \in \Substr(\FT)$, or
(ii) $X$ has an occurrence ending at a leaf of $\FT$.
Also, $X$ is said to be \emph{left-maximal} on $\FT$ if
(i) there are at least two distinct characters $a, b \in \Sigma$
    such that $aX, bX \in \Substr(\FT)$, or
(ii) $X$ has an occurrence beginning at the root of $\FT$.
Finally, $X$ is said to be \emph{maximal} on $\FT$ if $X$ is both right-maximal and
    left-maximal in $\FT$.
In the example of Figure~\ref{fig:FT_and_BT} (left),
$bc$ is left-maximal but is not right-maximal,
$ca$ is right-maximal but not left-maximal,
and $bca$ is maximal.

For any $X \in \Substr(\FT)$,
let $\Frmaximal(X)$, $\Flmaximal(X)$, and $\Fmaximal(X)$ respectively
denote the functions that map $X$ to the shortest right-maximal substring $X \beta$,
the shortest left-maximal substring $\alpha X$,
and the shortest maximal substring $\alpha X \beta$ that contain $X$ in $\FT$,
where $\alpha, \beta \in \Sigma^*$.

\subsection{Maximal Substrings on Backward Tries}
For any substring $Y$ in a backward trie $\BT$,
$Y$ is said to be \emph{left-maximal} on $\BT$ if
(i) there are at least two distinct characters $a, b \in \Sigma$
    such that $aY, bY \in \Substr(\BT)$, or
(ii) $Y$ has an occurrence beginning at a leaf of $\BT$.
Also, $Y$ is said to be \emph{right-maximal} on $\BT$ if
(i) there are at least two distinct characters $a, b \in \Sigma$
    such that $Ya, Yb \in \Substr(\BT)$, or
(ii) $Y$ has an occurrence ending at the root of $\BT$.
    Finally, $Y$ is said to be \emph{maximal} on $\BT$ if $Y$ is both right-maximal and
    left-maximal in $\BT$.
In the example of Figure~\ref{fig:FT_and_BT} (right),
$baaa$ is left-maximal but not right-maximal,
$aaa\$$ is right-maximal but not left-maximal,
and $baa$ is maximal.

For any $Y \in \Substr(\BT)$,
let $\Blmaximal(Y)$, $\Brmaximal(Y)$, and $\Bmaximal(Y)$ respectively
denote the functions that map $Y$ to the shortest left-maximal substring $\gamma Y$,
the shortest right-maximal substring $Y \delta$, and the shortest maximal substring
$\gamma Y \delta$ that contain $Y$ in $\BT$, where $\gamma, \delta \in \Sigma^*$.

Clearly, the afore-mentioned notions are symmetric over $\FT$ and $\BT$, namely:
\begin{fact} \label{fac:symmetry_tries}
String $X$ is right-maximal (resp. left-maximal) on $\FT$
iff $\rev{X}$ is left-maximal (resp. right-maximal) on $\BT$.
Also, $X$ is maximal on $\FT$ iff $\rev{X}$ is maximal on $\BT$.
\end{fact}

\section{Indexing Structures for Forward/Backward Tries}
\label{sec:indexing_structure_definitions}

In this section, we give formal definitions of
indexing structures for forward/backward tries.

A compact tree for a set $\mathbf{S}$ of strings
is a rooted tree such that 
(1) each edge is labeled by a non-empty substring of a string in $\mathbf{S}$,
(2) each internal node is branching, 
(3) the string labels of the out-going edges of each node begin with mutually distinct characters, and
(4) there is a path from the root that spells out each string in $\mathbf{S}$,
which may end on an edge.
Each edge of a compact tree is denoted by a triple $(u, \alpha, v)$
with $\alpha \in \Sigma^+$.
We call internal nodes that are branching as \emph{explicit nodes},
and we call loci that are on edges as \emph{implicit nodes}.
We will sometimes identify nodes with the substrings
that the nodes represent.

In what follows, we will consider DAG or tree data structures
built on a forward trie or backward trie.
For any DAG or tree data structure $\mathsf{D}$,
let $\NumNode{\mathsf{D}}$ and $\NumEdge{\mathsf{D}}$ denote
the numbers of nodes and edges in $\mathsf{D}$, respectively.

\subsection{Suffix Trees for Forward Tries}
\begin{figure}[ht!]
  \centering
         \includegraphics[scale=0.6]{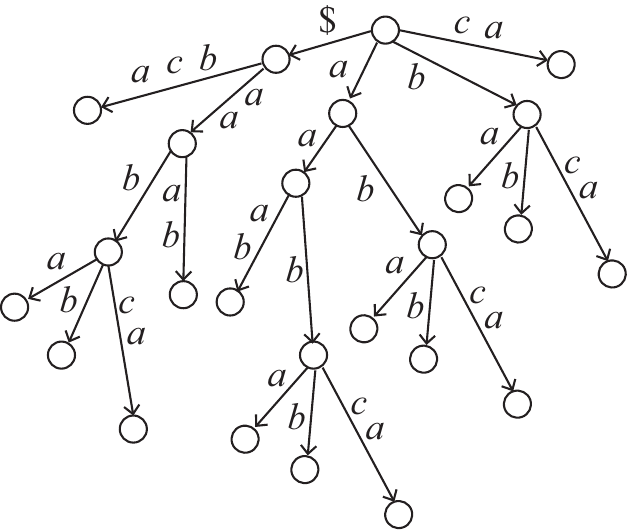}
	 \caption{$\STree(\FT)$ for the forward trie $\FT$ of Figure~\ref{fig:FT_and_BT}.}
         \label{fig:suffixtree_FT}
\end{figure}

The \emph{suffix tree} of a forward trie $\FT$,
denoted $\STree(\FT)$,
is a compact tree which represents $\Suffix(\FT)$.
See Figure~\ref{fig:suffixtree_FT}  for an example.

The nodes of $\STree(\FT)$ of Figure~\ref{fig:suffixtree_FT}
represent the right-maximal substrings in $\FT$ of Figure~\ref{fig:FT_and_BT},
e.g., $aab$ is right-maximal since it is immediately followed by
$a$, $b$, $c$ and also it ends at a leaf in $\FT$.
Hence $aab$ is a node in $\STree(\FT)$.
On the other hand, $aabc$ is not right-maximal
since it is immediately followed only by $c$ and hence it is not a node $\STree(\FT)$.

\begin{figure}[h!]
        \centering
        \includegraphics[scale=0.5]{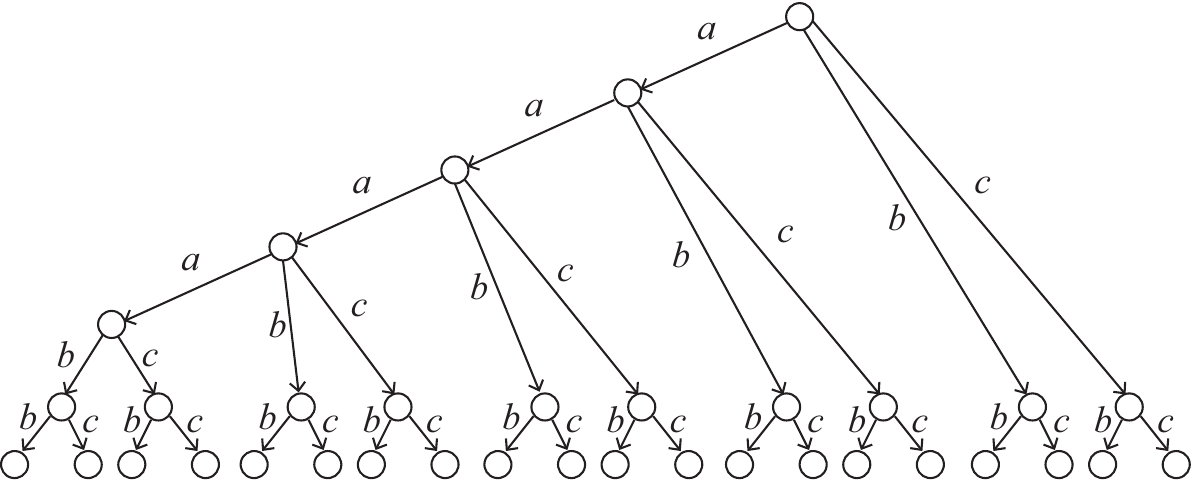}
	\caption{
          $\STree(\FT)$ for the forward trie $\FT$ of Figure~\ref{fig:FT_worstcase},
          which contains $k(k+1) = \Omega(n^2)$ nodes and edges where
          $n$ is the size of this $\FT$.
          In the example of Figure~\ref{fig:FT_worstcase}, $k = 4$
          and hence $\STree(\FT)$ here has $4\cdot5 = 20$ leaves.
          It is easy to modify the instance to a binary alphabet
          so that the suffix tree still has $\Omega(n^2)$ nodes.
	}
	\label{fig:ST_FT_worstcase}
\end{figure}

All non-root nodes in $\STree(\FT)$
represent right-maximal substrings on $\FT$.
Since now all internal nodes are branching,
and since there are at most $|\Suffix(\FT)|$ leaves, 
both the numbers of nodes and edges in $\STree(\FT)$ are proportional to
the number of suffixes in $\Suffix(\FT)$.
The following (folklore) quadratic bounds
hold due to Fact~\ref{fact:FT_BT_subst_suffix}-(b).
\begin{theorem} \label{theo:ST_FT}
  For any forward trie $\FT$ with $n$ nodes,
  $\NumNode{\STree(\FT)} = O(n^2)$ and
  $\NumEdge{\STree(\FT)}$ $= O(n^2)$.
  These upper bounds hold for any alphabet.
  For some forward trie $\FT$ with $n$ nodes,
  $\NumNode{\STree(\FT)}  = \Omega(n^2)$ and
  $\NumEdge{\STree(\FT)} = \Omega(n^2)$.
  These lower bounds hold for a constant-size or larger alphabet.
\end{theorem}
Figure~\ref{fig:ST_FT_worstcase} 
shows an example of the lower bounds of Theorem~\ref{theo:ST_FT}.

\subsection{Suffix Trees for Backward Tries}
The \emph{suffix tree} of a backward trie $\BT$,
denoted $\STree(\BT)$,
is a compact tree which represents $\Suffix(\BT)$.

\begin{figure}[h]
        \centering
         \includegraphics[scale=0.6]{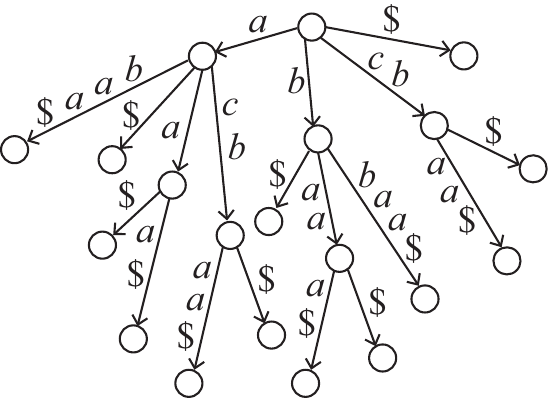}
	 \caption{$\STree(\BT)$ for the backward trie $\BT$ of Figure~\ref{fig:FT_and_BT}.}
         \label{fig:suffixtree_BT}
\end{figure}

See Figure~\ref{fig:suffixtree_BT}  for an example.

The nodes of $\STree(\BT)$ in Figure~\ref{fig:suffixtree_BT}
represent the right-maximal substrings in $\BT$ of Figure~\ref{fig:FT_and_BT},
e.g., $acb$ is right-maximal since it is immediately followed by
$a$ and $\$$. Hence $acb$ is a node in $\STree(\BT)$.
On the other hand, $ac$ is not right-maximal
since it is immediately followed only by $c$ and hence it is not a node $\STree(\BT)$.

Since $\STree(\BT)$ contains at most $n-1$ leaves by Fact~\ref{fact:FT_BT_subst_suffix}-(c)
and all internal nodes of $\Suffix(\BT)$ are branching,
the following precise bounds follow from Fact~\ref{fact:FT_BT_subst_suffix}-(c),
which were implicit in the literature~\cite{Kosaraju89a,breslauer_suffix_tree_tree_1998}.
\begin{theorem} \label{theo:ST_BT}
  For any backward trie $\BT$ with $n \geq 3$ nodes,
  $\NumNode{\STree(\BT)} \leq 2n-3$ and $\NumEdge{\STree(\BT)} \leq 2n - 4$,
  independently of the alphabet size.
\end{theorem}
The above bounds are tight since the theorem translates to
the suffix tree with $2m-1$ nodes and $2m-2$ edges for a string of length $m$
(e.g., $a^{m-1}b$), which can be represented as a path tree with $n = m+1$ nodes.
By representing each edge label $\alpha$ by a pair $\langle v, u \rangle$ of nodes
in $\BT$ such that $\alpha = \Bpathstr(u, v)$,
$\STree(\BT)$ can be stored with $O(n)$ space.


\vspace{1pc}
\noindent \textbf{Suffix Links and Weiner Links:}
For each explicit node $aU$ of the suffix tree $\STree(\BT)$
of a backward trie $\BT$
with $a \in \Sigma$ and $U \in \Sigma^*$, let $\suflink(aU) = U$.
This is called the \emph{suffix link} of node $aU$.
For each explicit node $V$ and $a \in \Sigma$,
we also define the \emph{reversed suffix link} $\W{a}{V} = aVX$
where $X \in \Sigma^*$ is the shortest string
such that $aVX$ is an explicit node of $\STree(\BT)$.
$\W{a}{V}$ is undefined if $aV \notin \Substr(\BT)$.
These reversed suffix links are also called as \emph{Weiner links}
(or \emph{W-link} in short)~\cite{BreslauerI13}.
A W-link $\W{a}{V} = aVX$ is said to be \emph{hard}
if $X = \varepsilon$, and \emph{soft} if $X \in \Sigma^+$.
%
The suffix links, hard and soft W-links of nodes in
the suffix tree $\STree(\FT)$ of a forward trie $\FT$
are defined analogously.


\vspace{1pc}
\noindent \textbf{Suffix Arrays of Forward/Backward Tries}

An array representation of the list of the leaves of
$\STree(\FT)$ (resp. $\STree(\BT)$)
sorted in the lexicographical order is the \emph{suffix array}
of the forward trie $\FT$ (resp. backward trie $\BT$).
Hence, the following corollaries are immediate from Theorem~\ref{theo:ST_FT}
and Theorem~\ref{theo:ST_BT}, respectively:

\begin{corollary}
  For any forward trie $\FT$ with $n$ nodes,
  the size of the suffix array for $\FT$ is $O(n^2)$.
  This upper bound holds for any alphabet.
  For some forward trie $\FT$ with $n$ nodes,
  the size of the suffix array for $\FT$ is $\Omega(n^2)$.
  This lower bound holds for a constant-size or larger alphabet.
\end{corollary}

\begin{corollary}[\cite{FerraginaLMM09}]
  For any backward trie $\BT$ with $n$ nodes,
  the size of the suffix array is exactly $n-1$.
\end{corollary}

\subsection{DAWGs for Forward Tries}
The \emph{directed acyclic word graph}
(\emph{DAWG}) of a forward trie $\FT$ is a (partial) DFA
that recognizes all substrings in $\Substr(\FT)$.
Hence, 
the label of every edge of $\DAWG(\FT)$ is a single character from $\Sigma$.
$\DAWG(\FT)$ is formally defined as follows:
For any substring $X$ from $\Substr(\FT)$, 
let $\FEeqc{X}$ denote the equivalence class w.r.t. $\Flmaximal(X)$.
There is a one-to-one correspondence between 
the nodes of $\DAWG(\FT)$ 
and the equivalence classes $\FEeqc{\cdot}$,
and hence we will identify the nodes of $\DAWG(\FT)$
with their corresponding equivalence classes $\FEeqc{\cdot}$.

\begin{figure}[htb!]
        \centering
          \includegraphics[scale=0.6]{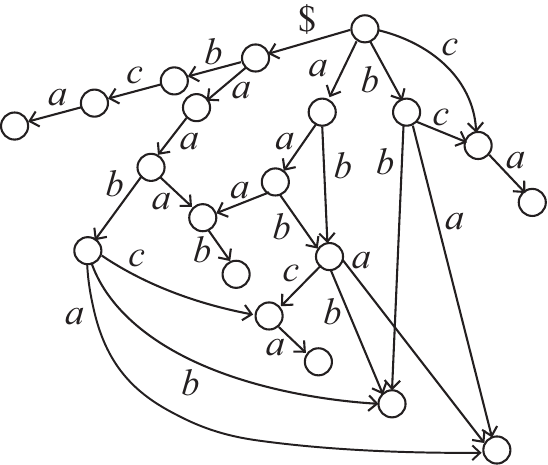}
          \caption{$\DAWG(\FT)$ for the forward trie $\FT$ of Figure~\ref{fig:FT_and_BT}.}
          \label{fig:DAWG_FT}
\end{figure}

See Figure~\ref{fig:DAWG_FT} for an example.

The nodes of $\DAWG(\FT)$ of Figure~\ref{fig:DAWG_FT}
represent the equivalence classes w.r.t. 
the left-maximal substrings in $\FT$ of Figure~\ref{fig:FT_and_BT},
e.g., $aab$ is left-maximal since it is immediately followed by
$a$ and $\$$ and hence it is the longest string in the node that
represents $aab$. This node also represents the suffix $ab$ of $aab$,
since $\Flmaximal(ab) = aab$.

By the definition of equivalence classes,
every member of $\FEeqc{X}$ is a suffix of $\Flmaximal(X)$.
If $X, Xa$ are substrings in $\Substr(\FT)$ and $a \in \Sigma$,
then there exists an edge labeled with character $a \in \Sigma$
from node $\FEeqc{X}$ to node $\FEeqc{Xa}$ in $\DAWG(\FT)$.
This edge is called \emph{primary} if $|\Flmaximal(X)| + 1 = |\Flmaximal(Xa)|$,
and is called \emph{secondary} otherwise.
For each node $\FEeqc{X}$ of $\DAWG(\FT)$ with $|X| \geq 1$,
let $\suflink(\FEeqc{X}) = Z$, where
$Z$ is the longest suffix of $\Flmaximal(X)$ not belonging to $\FEeqc{X}$.
This is the \emph{suffix link} of this node $\FEeqc{X}$.
%

%

Mohri et al.~\cite{MohriMW09} introduced the \emph{suffix automaton} for an acyclic
DFA $\mathsf{G}$,
which is a small DFA that represents all suffixes of strings accepted by $\mathsf{G}$.
They considered equivalence relation $\equiv$
of substrings $X$ and $Y$ in an acyclic DFA $\mathsf{G}$
such that $X \equiv Y$ iff the following paths of the occurrences of $X$ and $Y$
in $\mathsf{G}$ are equal.
Mohri et al.'s equivalence class is identical to our equivalence class $\FEeqc{X}$
when $\mathsf{G} = \FT$.
To see why, recall that $\Flmaximal(X) = \alpha X$ is the shortest substring of $\FT$
such that $\alpha X$ is left-maximal, where $\alpha \in \Sigma^*$.
Therefore, $X$ is a suffix of $\Flmaximal(X)$ and
the following paths of the occurrences of $X$ in $\FT$ are identical to
the following paths of the occurrences of $\Flmaximal(X)$ in $\FT$.
Hence, in the case where the input DFA $\mathsf{G}$ is in form of a forward trie $\FT$
such that its leaves are the accepting states,
then Mohri et al.'s suffix automaton is identical to our DAWG for $\FT$.
Mohri et al.~\cite{MohriMW09} showed the following:

\begin{theorem}[Corollary 2 of~\cite{MohriMW09}] \label{theo:DAWG_FT_node}
  For any forward trie $\FT$ with $n \geq 3$ nodes,
  $\NumNode{\DAWG(\FT)} \allowbreak \leq 2n - 3$, independently of the alphabet size.
\end{theorem}
We remark that Theorem~\ref{theo:DAWG_FT_node}
is immediate from Theorem~\ref{theo:ST_BT} and Fact~\ref{fac:symmetry_tries}.
This is because there is a one-to-one correspondence between the nodes
of $\DAWG(\FT)$ and the nodes of $\STree(\BT)$,
which means that $\NumNode{\DAWG(\FT)} = \NumNode{\STree(\BT)}$.
Recall that the bound in Theorem~\ref{theo:DAWG_FT_node} is only
on the number of \emph{nodes} in $\DAWG(\FT)$.
We shall show later that the number of \emph{edges} in $\DAWG(\FT)$ is
$\Omega(\sigma n)$ in the worst case,
which can be $\Omega(n^2)$ for a large alphabet.

\subsection{DAWGs for Backward Tries}
The DAWG of a backward trie $\BT$, denoted $\DAWG(\BT)$, is
a (partial) DFA that recognizes all strings in $\Substr(\BT)$.
The label of every edge of $\DAWG(\BT)$ is a single character from $\Sigma$.
$\DAWG(\BT)$ is formally defined as follows:
For any substring $Y$ from $\Substr(\BT)$, 
let $\BEeqc{Y}$ denote the equivalence class w.r.t. $\Blmaximal(Y)$.
There is a one-to-one correspondence between 
the nodes of $\DAWG(\BT)$ 
and the equivalence classes $\BEeqc{\cdot}$,
and hence we will identify the nodes of $\DAWG(\BT)$
with their corresponding equivalence classes $\BEeqc{\cdot}$.

\begin{figure}[h]
        \centering
          \includegraphics[scale=0.6]{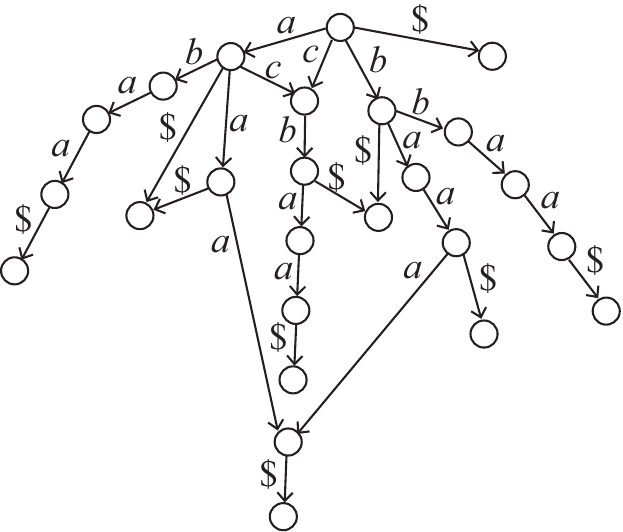}
          \caption{$\DAWG(\BT)$ for the backward trie $\BT$ of Figure~\ref{fig:FT_and_BT}.}
          \label{fig:DAWG_BT}
\end{figure}

See Figure~\ref{fig:DAWG_BT} for an example.

The nodes of $\DAWG(\BT)$ in Figure~\ref{fig:DAWG_BT} represent the equivalence classes w.r.t. 
the left-maximal substrings in $\BT$ of Figure~\ref{fig:FT_and_BT},
e.g., $ac$ is left-maximal since it begins at a leaf in $\BT$,
and hence it is the longest string in the node that
represents $ac$. This node also represents the suffix $c$ of $ac$,
since $\Blmaximal(c) = ac$.

The notions of primary edges, secondary edges,
and the suffix links of $\DAWG(\BT)$ are defined
in a similar manner to $\DAWG(\FT)$, by using the equivalence classes $\BEeqc{Y}$
for substrings $Y$ in the backward trie $\BT$.

\vspace*{1pc}
\noindent \textbf{Symmetries between Suffix Trees and DAWGs:}
The well-known \emph{symmetry} between the suffix trees and the DAWGs
(refer to~\cite{blumer85:_small_autom_recog_subwor_text,Blumer87,cr:94})
also holds in our case of forward and backward tries.
Namely, the suffix links of $\DAWG(\FT)$ (resp. $\DAWG(\BT)$) are 
the (reversed) edges of $\STree(\BT)$ (resp. $\STree(\FT)$).
Also, the hard W-links of $\STree(\FT)$ (resp. $\STree(\BT)$)
are the primary edges of $\DAWG(\BT)$ (resp. $\DAWG(\FT)$),
and the soft W-links of $\STree(\FT)$ (resp. $\STree(\BT)$)
are the secondary edges of $\DAWG(\BT)$ (resp. $\DAWG(\FT)$).


\subsection{CDAWGs for Forward Tries}
The \emph{compact directed acyclic word graph} (\emph{CDAWG}) of a forward trie $\FT$,
denoted $\CDAWG(\FT)$, is
the edge-labeled DAG where the nodes
correspond to the equivalence class of $\Substr(\FT)$ w.r.t. $\Fmaximal(\cdot)$.
In other words, $\CDAWG(\FT)$ can be obtained
by merging isomorphic subtrees of $\STree(\FT)$ rooted at internal nodes
and merging leaves that are equivalent under $\Fmaximal(\cdot)$, or
by contracting non-branching paths of $\DAWG(\FT)$.

\begin{figure}[thb!]
  \centering
          \includegraphics[scale=0.6]{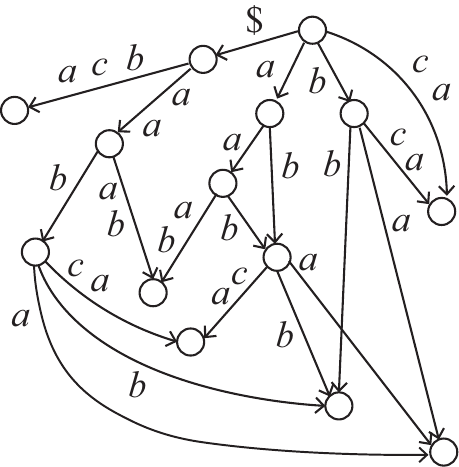}
          \caption{$\CDAWG(\FT)$ for the forward trie $\FT$ of Figure~\ref{fig:FT_and_BT}.}
          \label{fig:CDAWG_FT}
\end{figure}

See Figure~\ref{fig:CDAWG_FT} for an example.

The nodes of $\CDAWG(\FT)$ of Figure~\ref{fig:CDAWG_FT} represent
the equivalence classes w.r.t.
the maximal substrings in $\FT$ of Figure~\ref{fig:FT_and_BT},
e.g., $aab$ is maximal since it is both left- and right-maximal as described above
and hence it is the longest string in the node that represents $aab$.
This node also represents the suffix $ab$ of $aab$,
since $\Fmaximal(ab) = aab$.

\begin{theorem}[\cite{InenagaHSTA01}] \label{theo:CDAWG_FT_constant}
  For any forward trie $\FT$ with $n$ nodes over a \emph{constant-size alphabet},\\
  $\NumNode{\CDAWG(\FT)} = O(n)$ and $\NumEdge{\CDAWG(\FT)} = O(n)$.
\end{theorem}
We emphasize that the above result by Inenaga et al.~\cite{InenagaHSTA01}
states size bounds of $\CDAWG(\FT)$ only in the case where $\sigma = O(1)$.
We will later show that this bound does not hold
for the number of edges, in the case of a large alphabet.

\subsection{CDAWGs for Backward Tries}
The \emph{compact directed acyclic word graph}
(\emph{CDAWG}) of a backward trie $\BT$,
denoted $\CDAWG(\BT)$, is
the edge-labeled DAG where the nodes
correspond to the equivalence class of $\Substr(\BT)$ w.r.t. $\Bmaximal(\cdot)$.
Similarly to its forward trie counterpart,
$\CDAWG(\BT)$ can be obtained
by merging isomorphic subtrees of $\STree(\BT)$ rooted at internal nodes
and merging leaves that are equivalent under $\Fmaximal(\cdot)$, or
by contracting non-branching paths of $\DAWG(\BT)$.

\begin{figure}[h]
        \centering
          \includegraphics[scale=0.6]{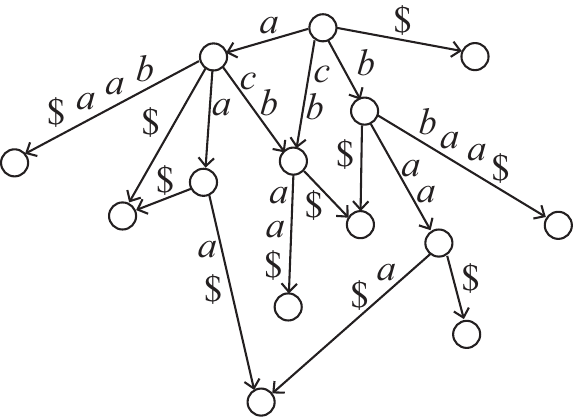}
          \caption{$\CDAWG(\BT)$ for the backward trie $\BT$ of Figure~\ref{fig:FT_and_BT}.}
          \label{fig:CDAWG_BT}
\end{figure}

See Figure~\ref{fig:CDAWG_BT} for an example.

The nodes of $\CDAWG(\BT)$ in Figure~\ref{fig:CDAWG_BT} represent
the equivalence classes w.r.t.
the maximal substrings in $\FT$ of Figure~\ref{fig:FT_and_BT},
e.g., $acb$ is maximal since it is both left- and right-maximal in $\BT$
and hence it is the longest string in the node that represents $acb$.
This node also represents the suffix $cb$ of $acb$,
since $\Fmaximal(cb) = acb$.
Notice that there is a one-to-one correspondence
between the nodes of $\CDAWG(\FT)$ in Figure~\ref{fig:CDAWG_BT}
and the nodes of $\CDAWG(\BT)$ in Figure~\ref{fig:CDAWG_FT}.
In other words, $X$ is the longest string represented by a node in $\CDAWG(\FT)$
iff $Y = \rev{X}$ is the longest string represented by a node in $\CDAWG(\BT)$.
For instance, $aab$ is the longest string represented by a node of $\CDAWG(\FT)$
and $baa$ is the longest string represented by a node of $\CDAWG(\BT)$,
and so on.
Hence the numbers of nodes in $\CDAWG(\FT)$ and $\CDAWG(\BT)$ are equal.

\section{New Size Bounds on Indexing Forward/\\Backward Tries}
\label{sec:new_bounds}

To make the analysis simpler, we assume
each of the roots, the one of $\FT$ and the corresponding one of $\BT$,
is connected to an auxiliary node $\bot$ with an edge labeled
by a unique character $\$$ that does not appear elsewhere in $\FT$ or in $\BT$.

\subsection{Size Bounds for DAWGs for Backward Tries}

We begin with the size bounds for the DAWG for a backward trie.
\begin{theorem} \label{theo:DAWG_BT}
  For any backward trie $\BT$ with $n$ nodes,
  $\NumNode{\DAWG(\BT)} = O(n^2)$ and \\
  $\NumEdge{\DAWG(\BT)} = O(n^2)$.
  These upper bounds hold for any alphabet.
  For some backward trie $\BT$ with $n$ nodes,
  $\NumNode{\DAWG(\BT)} = \Omega(n^2)$ and 
  $\NumEdge{\DAWG(\BT)}$ $= \Omega(n^2)$. 
  These lower bounds hold for a constant-size or larger alphabet.
\end{theorem}

\begin{proof}
  The bounds $\NumNode{\DAWG(\BT)} = O(n^2)$ and $\NumNode{\DAWG(\BT)} = \Omega(n^2)$
  for the number of nodes 
  immediately follow from Fact~\ref{fac:symmetry_tries} and Theorem~\ref{theo:ST_FT}.

  Since each internal node in $\DAWG(\BT)$ has at least one out-going edge
  and since $\NumNode{\DAWG(\BT)}$ $= \Omega(n^2)$,
  the lower bound $\NumEdge{\DAWG(\BT)} = \Omega(n^2)$ for the number of edges is immediate.
  To show the upper bound for the number of edges,
  we consider the \emph{suffix trie} of $\BT$.
  Since there are $O(n^2)$ pairs of nodes in $\BT$,
  the number of substrings in $\BT$ is clearly $O(n^2)$.
  Thus, the numbers of nodes and edges in the suffix trie of $\BT$ are $O(n^2)$.
  Hence $\NumEdge{\DAWG(\BT)} = O(n^2)$.
\end{proof}

\subsection{Size Bounds for DAWGs for Forward Tries}

In this subsection, we consider the size bounds for the DAWG of a forward trie.
\begin{theorem} \label{theo:DAWG_FT_edge}
  For any forward trie $\FT$ with $n$ nodes,
  $\NumEdge{\DAWG(\FT)} = O(\sigma n)$.
  For some forward trie $\FT$ with $n$ nodes,
  $\NumEdge{\DAWG(\FT)} = \Omega(\sigma n)$ 
  which is $\Omega(n^2)$ for a large alphabet of size $\sigma = \Theta(n)$.
\end{theorem}

\begin{figure}[htb]
        \centering
	\raisebox{3mm}{\includegraphics[scale=0.6]{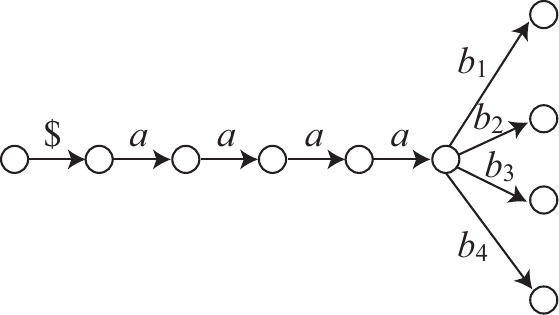}}
        \hfill
        \includegraphics[scale=0.6]{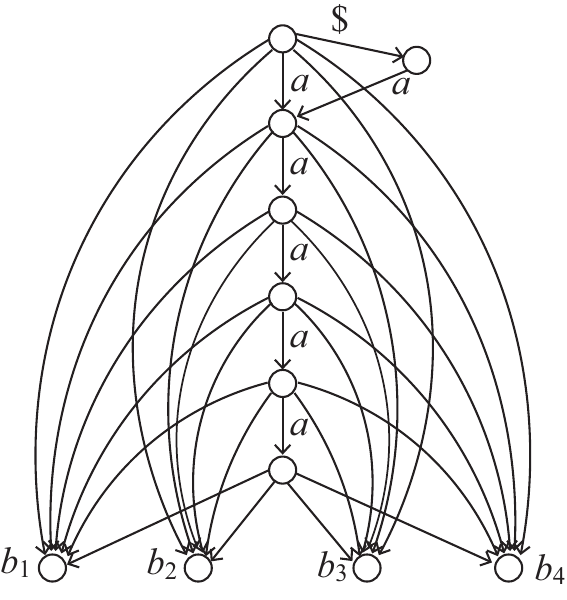}
	\caption{
          Left: The broom-like $\FT$ for the lower bound of Theorem~\ref{theo:DAWG_FT_edge},
          where $n = 10$ and $\sigma = (n-2)/2 = 4$.
          Right: $\DAWG(\FT)$ for this $\FT$ has $\Omega(n^2)$ edges. The labels $b_1, \ldots, b_4$ of the in-coming edges to the sinks are omitted for better visualization.
	}
	\label{fig:DAWG_FT_edge_lowerbound}
\end{figure}

\begin{proof}
  Since each node of $\DAWG(\FT)$ can have at most $\sigma$
  out-going edges, $\NumEdge{\DAWG(\FT)} = O(\sigma n)$ follows from
  Theorem~\ref{theo:DAWG_FT_node}.

  To obtain the lower bound $\NumEdge{\DAWG(\FT)} = \Omega(\sigma n)$,
  we consider $\FT$ which has a broom-like shape
  such that there is a single path of length $n - \sigma - 1$ from the root to a node $v$
  which has out-going edges 
  with $\sigma$ distinct characters $b_1, \ldots, b_\sigma$
  (see Figure~\ref{fig:DAWG_FT_edge_lowerbound} for illustration.)
  Since the root of $\FT$ is connected with the auxiliary node $\bot$ with an edge labeled $\$$,
  each root-to-leaf path in $\FT$ represents $\$ a^{n - \sigma + 1} b_i$ for $1 \leq i \leq \sigma$.
  Now $a^k$ for each $1 \leq k \leq n - \sigma -2$ is left-maximal
  since it is immediately preceded by $a$ and $\$$.
  Thus $\DAWG(\FT)$ has at least $n - \sigma - 2$ internal nodes, each representing
  $a^k$ for $1 \leq k \leq n - \sigma - 2$.
  On the other hand, each $a^k \in \Substr(\FT)$ is immediately followed by
  $b_i$ with all $1 \leq i \leq \sigma$.
  Hence, $\DAWG(\FT)$ contains $\sigma (n-\sigma-2) = \Omega(\sigma n)$ edges
  when $n - \sigma - 2 = \Omega(n)$.
  By choosing e.g. $\sigma \approx n/2$, we obtain $\DAWG(\FT)$ that contains $\Omega(n^2)$ edges. 
\end{proof}

\subsubsection{On Construction of DAWGs for Forward Tries}

Mohri et al. (Proposition 4 of~\cite{MohriMW09})
claimed that one can construct $\DAWG(\FT)$ in time proportional to its size.
The following corollary is immediate from Theorem~\ref{theo:DAWG_FT_edge}:
\begin{corollary}
  The DAWG construction algorithm of~\cite{MohriMW09}
  applied to a forward trie with $n$ nodes
  must take at least $\Omega(n^2)$ time in the worst case
  for an alphabet of size $\sigma = \Theta(n)$.
\end{corollary}
Mohri et al.'s proof for Proposition 4 in~\cite{MohriMW09} contains yet another issue:
They claimed that the number of redirections of secondary edges
during the construction of $\DAWG(\FT)$ can be bounded by the number $n$ of nodes in $\FT$,
but this is not true.
Breslauer~\cite{breslauer_suffix_tree_tree_1998} already pointed out this issue
in his construction for $\STree(\BT)$ that is based on Weiner's algorithm
(recall that Weiner's suffix tree construction algorithm inherently builds
the DAWG for the reversed strings),
and he overcame this difficulty by using $\sigma$ nearest marked ancestor data structures
for all $\sigma$ characters, instead of explicitly maintaining soft W-links.
This leads to $O(\sigma n)$-time and space construction
for $\STree(\BT)$ that works in $O(n)$ time and space for constant-size alphabets.
In Section~\ref{sec:DAWG_FT_construction}
we will present how to build an $O(n)$-space \emph{implicit}
representation of $\DAWG(\FT)$ in $O(n)$ time and working space
for larger alphabets of size $\sigma = O(n)$.

\subsubsection{Size Bounds of Smallest Automaton for Forward Tries}

Consider the smallest DFA that accepts the set $\Suffix(\FT)$
of suffixes of forward trie $\FT$.
It is apparent that $\DAWG(\FT)$ is not the smallest such DFA,
since it can have multiple sinks as in the example of Figure~\ref{fig:DAWG_FT}.
However, our lower bound instance for $\DAWG(\FT)$ also gives
$\Omega(\sigma n)$ transitions in the smallest DFA.
See Figure~\ref{fig:DAWG_FT_edge_lowerbound}.
The smallest DFA that accepts the same language as the DAWG in Figure~\ref{fig:DAWG_FT_edge_lowerbound}
is the one obtained by merging all the sink nodes which correspond to the final states.
However, this does not reduce the number of edges (transitions) at all.
Hence, this smallest DFA still has $\Omega(\sigma n)$ transitions.
\begin{theorem}
  For any forward trie $\FT$ with $n$ nodes,
  the number of transitions in the smallest DFA
  that accepts $\Suffix(\FT)$ is $O(\sigma n)$.
  For some forward trie $\FT$ with $n$ nodes,
  the number of transitions in the smallest DFA
  that accepts $\Suffix(\FT)$ is $\Omega(\sigma n)$,
  which is $\Omega(n^2)$ for a large alphabet of size $\sigma = \Theta(n)$.
\end{theorem}
The same bounds hold for the smallest DFA that accepts
the set $\Substr(\FT)$ of substrings in forward trie $\FT$.

\subsection{Size Bounds for CDAWGs for Backward Tries}

We begin with the size bounds of the CDAWG for a backward trie.

\begin{figure}[htb]
        \centerline{
          \includegraphics[scale=0.6]{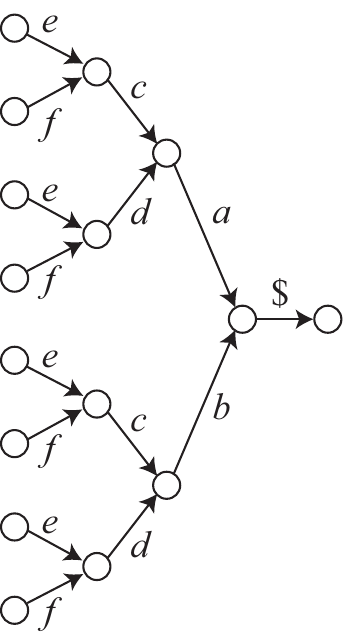}
          \hfil
          \includegraphics[scale=0.6]{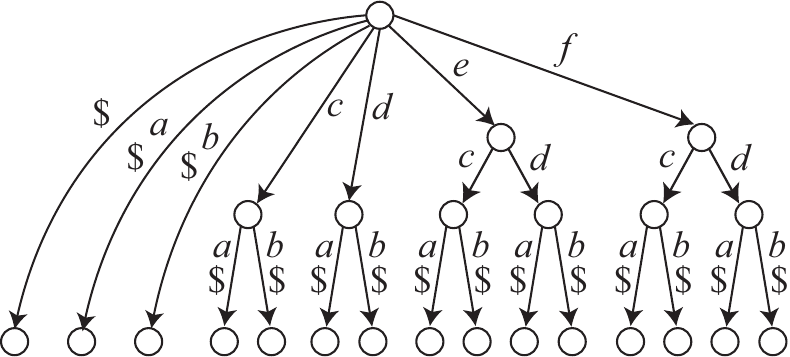}
        }
	\caption{Left: A backward trie which gives the largest number of nodes and edges in the CDAWG for backward tries. Here, the sub-alphabets are $\{a, b\}$ for depth 2, $\{c, d\}$ for depth 3, and $\{e, f\}$ for depth 4. Right: The CDAWG for the backward trie. Notice that no isomorphic subtrees are merged under our definition of equivalence classes. For instance, consider substrings $c$ and $d$. Since $\Bmaximal(c) = \Brmaximal(\Blmaximal(c)) = \Brmaximal(c) = c \neq d = \Brmaximal(\Blmaximal(d)) = \Brmaximal(d) = \Bmaximal(d)$, the isomorphic subtrees rooted at $c$ and $d$ are not merged. By the same reasoning, isomorphic subtrees (including sink nodes) are not merged in this CDAWG.}
	\label{fig:CDAWG_BT_worstcase}
\end{figure}

\begin{theorem} \label{theo:CDAWG_BT}
  For any backward trie $\BT$ with $n$ nodes,
  $\NumNode{\CDAWG(\BT)} \leq 2n - 3$ and \\
  $\NumEdge{\CDAWG(\BT)} \leq 2n -4$.
  These bounds are independent of the alphabet size.
\end{theorem}
\begin{proof}
  Since any maximal substring in $\Substr(\BT)$ is right-maximal in $\Substr(\BT)$,
  by Theorem~\ref{theo:ST_BT} we have
  $\NumNode{\CDAWG(\BT)} \leq \NumNode{\STree(\BT)} \leq 2n - 3$
  and $\NumEdge{\CDAWG(\BT)} \leq \NumEdge{\STree(\BT)} \leq 2n - 4$.
\end{proof}
The bounds in Theorem~\ref{theo:CDAWG_BT} are tight:
Consider an alphabet
$\{a_1, \ldots, a_{\lceil \log_2 n \rceil}, b_1, \ldots, b_{\lceil \log_2 n \rceil}, \$\}$
of size $2 \lceil \log_2 n \rceil + 1$
and a binary backward trie $\BT$ with $n$ nodes
where the binary edges at each depth $d \geq 2$ are labeled by the sub-alphabet $\{a_d, b_d\}$ of size $2$
(see also Figure~\ref{fig:CDAWG_BT_worstcase}).
Because every suffix $S \in \Suffix(\BT)$ is maximal in $\BT$,
$\CDAWG(\BT)$ for this $\BT$ contains $n-1$ sinks.
Also, since for each suffix $S$ in $\BT$ there is a unique suffix
$S' \neq S$ that shares the longest common prefix with $S$,
$\CDAWG(\BT)$ for this $\BT$ contains $n-2$ internal nodes (including the source).
This also means $\CDAWG(\BT)$ is identical to $\STree(\BT)$ for this backward trie $\BT$.

\subsection{Size Bounds for CDAWGs for Forward Tries}

Next, we turn our attention to the size bounds of the CDAWG for a forward trie.

\begin{theorem} \label{theo:CDAWG_FT}
  For any forward trie $\FT$ with $n$ nodes,
  $\NumNode{\CDAWG(\FT)} \leq 2n - 3$ and \\
  $\NumEdge{\CDAWG(\FT)} = O(\sigma n)$.
  For some forward trie $\FT$ with $n$ nodes,
  $\NumEdge{\CDAWG(\FT)} = \Omega(\sigma n)$ 
  which is $\Omega(n^2)$ for a large alphabet of size $\sigma = \Theta(n)$.
\end{theorem}

\begin{proof}
  It immediately follows from
  Fact~\ref{fact:FT_BT_subst_suffix}-(a), Fact~\ref{fac:symmetry_tries},
  and Theorem~\ref{theo:CDAWG_BT}
  that $\NumNode{\CDAWG(\FT)}$ $= \NumNode{\CDAWG(\BT)} \leq 2n - 3$.
  Since a node in $\CDAWG(\FT)$ can have
  at most $\sigma$ out-going edges,
  the upper bound $\NumEdge{\CDAWG(\FT)} = O(\sigma n)$ of the number of edges trivially holds.
  To obtain the lower bound,
  we consider the same broom-like forward trie $\FT$
  as in Theorem~\ref{theo:DAWG_FT_edge}.
  In this $\FT$, $a^k$ for each $1 \leq k \leq n - \sigma -2$ is maximal
  and thus $\CDAWG(\FT)$ has at least $n - \sigma - 2$ internal nodes each representing
  $a^k$ for $1 \leq k \leq n - \sigma - 2$.
  By the same argument as Theorem~\ref{theo:DAWG_FT_edge},
  $\CDAWG(\FT)$ for this $\FT$ contains at least $\sigma(n-\sigma-2) = \Omega(\sigma n)$
  edges, which amounts to $\Omega(n^2)$ for a large alphabet of size e.g. $\sigma \approx n/2$.
\end{proof}
The upper bound of Theorem~\ref{theo:CDAWG_FT} generalizes the bound of Theorem~\ref{theo:CDAWG_FT_constant} for constant-size alphabets.
Remark that $\CDAWG(\FT)$ for the broom-like $\FT$
of Figure~\ref{fig:DAWG_FT_edge_lowerbound}
is almost identical to $\DAWG(\FT)$,
except for the unary path $\$a$ that is compacted in $\CDAWG(\FT)$.

\section{Constructing $O(n)$-size Representation of $\DAWG(\FT)$ in $O(n)$ time}
\label{sec:DAWG_FT_construction}

We have seen that $\DAWG(\FT)$ for any forward trie $\FT$ with $n$ nodes
contains only $O(n)$ nodes, but can have $\Omega(\sigma n)$ edges
for some $\FT$ over an alphabet of size $\sigma$
ranging from $\Theta(1)$ to $\Theta(n)$.
Thus some $\DAWG(\FT)$ can have $\Theta(n^2)$ edges for $\sigma = \Theta(n)$
(Theorem~\ref{theo:DAWG_FT_node} and Theorem~\ref{theo:DAWG_FT_edge}).
Hence, in general
it is impossible to build an \emph{explicit} representation of $\DAWG(\FT)$
within linear $O(n)$-space.
By an explicit representation we mean an implementation of
$\DAWG(\FT)$ where each edge is represented by a pointer between two nodes.

Still, we show that there exists an $O(n)$-space \emph{implicit} representation
of $\DAWG(\FT)$ for any alphabet of size
$\sigma$ ranging from $\Theta(1)$ to $\Theta(n)$,
that allows us $O(\log \sigma)$-time access 
to each edge of $\DAWG(\FT)$.
This is trivial in case $\sigma = O(1)$,
and hence in what follows we consider an alphabet of size
$\sigma$ such that $\sigma$ ranges from $\omega(1)$ to $\Theta(n)$.
Also, we suppose that our alphabet is 
an integer alphabet $\Sigma = [1..\sigma]$ of size $\sigma$.
Then, we show that such an implicit representation of $\DAWG(\FT)$
can be built in $O(n)$ time and working space.

Based on the property stated in Section~\ref{sec:indexing_structure_definitions},
constructing $\DAWG(\FT)$ reduces
to maintaining hard and soft W-links over $\STree(\BT)$.
Our data structure explicitly stores all $O(n)$ hard W-links,
while it only stores carefully selected $O(n)$ soft W-links.
The other soft W-links can be 
simulated by these explicitly stored W-links, in $O(\log \sigma)$ time each.

Our algorithm is built upon the following facts which are adapted from~\cite{FischerG2015,abs-1302-3347}:
\begin{fact} \label{fact:soft_wlink_monotonicity}
  Let $a$ be any character from $\Sigma$.
  \begin{enumerate}
  \item[(a)] If there is a (hard or soft) W-link $\W{a}{V}$ for a node $V$ in $\STree(\BT)$,
  then there is a (hard or soft) W-link $\W{a}{U}$ for any ancestor $U$ of $V$ in $\STree(\BT)$.

  \item[(b)] If two nodes $U$ and $V$ have hard W-links $\W{a}{U}$ and $\W{a}{V}$, then the lowest common ancestor (LCA) $Z$ of $U$ and $V$ also has a hard W-link $\W{a}{Z}$.
  \end{enumerate}
  In the following statements (c), (d), and (e),
  let $V$ be any node of $\STree(\BT)$ such that
  $V$ has a \emph{soft} W-link $\W{a}{V}$ for $a \in \Sigma$.
  \begin{enumerate}
  \item[(c)] There is a descendant $U$ of $V$
  s.t. $U \neq V$ and $U$ has a hard W-link $\W{a}{V}$.

  \item[(d)] The highest descendant of $V$ that has a hard W-link for character $a$ is unique. This fact follows from (b).

  \item[(e)] Let $Z$ be the unique highest descendant of $V$ that has a hard W-link $\W{a}{Z}$.
    For every node $X$ in the path from $V$ to $Z$,
    $\W{a}{X} = \W{a}{Z}$, i.e.
    the W-links of all nodes in this path for character $a$ point to the same node in $\STree(\BT)$.
  \end{enumerate}
\end{fact}

\subsection{Compact Representation of Weiner Links}

\begin{figure}[htb]
 \centering
 \includegraphics[scale=0.5]{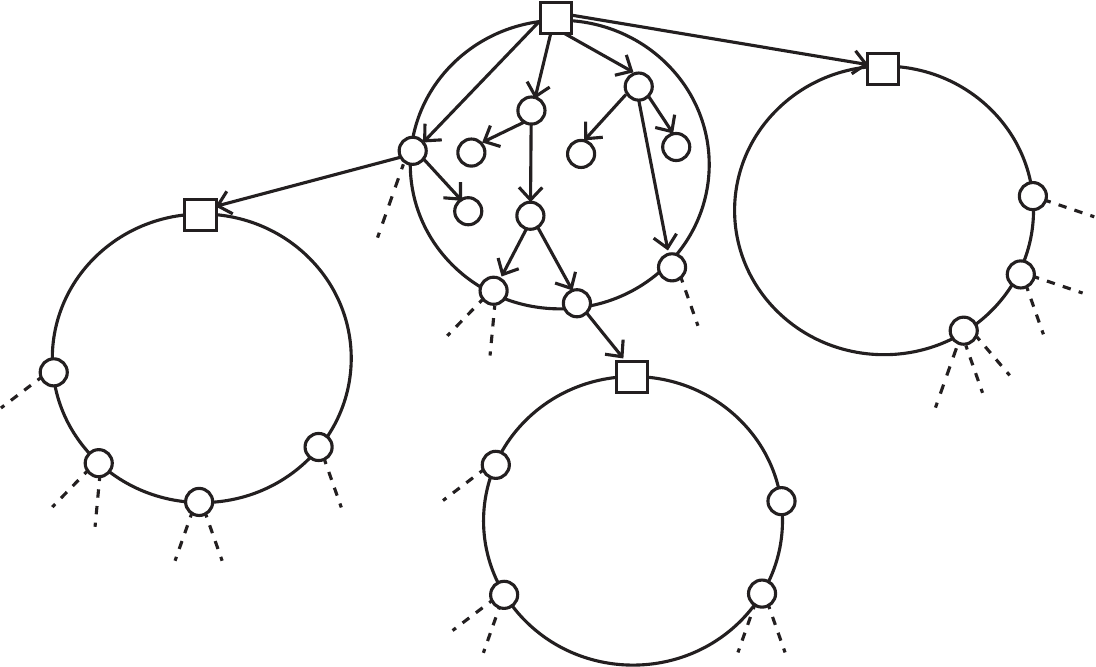}
 \caption{Illustration for our micro-macro tree decomposition of $\STree(\BT)$.
 The large circles represent micro tree of size $O(\sigma)$ each,
 and the square nodes are the roots of the micro trees.
 The macro tree is the induced tree from the square nodes.}
 \label{fig:micro-macro-tree-decomposition}
\end{figure}

We construct a micro-macro tree decomposition~\cite{AlstrupSS97} of $\STree(\BT)$ in a similar manner to~\cite{Gawrychowski14},
such that the nodes of $\STree(\BT)$ are partitioned into
$O(n / \sigma)$ connected components (called \emph{micro-trees}), each of which contains
$O(\sigma)$ nodes.
(see Figure~\ref{fig:micro-macro-tree-decomposition}).
Such a decomposition always exists and can be computed in $O(n)$ time.
The \emph{macro tree} is the induced tree from the roots of the micro trees,
and thus the macro tree contains $O(n / \sigma)$ nodes.

In every node $V$ of the macro tree,
we explicitly store all soft and hard W-links from $V$.
Since there can be at most $\sigma$ W-links from $V$,
this requires $O(n)$ total space for all nodes in the macro tree.
Let $\microtree$ denote any micro tree.
We compute the ranks of all nodes in a pre-order traversal in $\microtree$.
Let $a \in \Sigma$ be any character 
such that there is a node $V$ in $\microtree$ that has
a hard W-link $\W{a}{V}$.
Let $\mathsf{P}_a^{\microtree}$ denote an array that stores
a sorted list of pre-order ranks of nodes $V$ in $\microtree$
that have hard W-links for character $a$.
Hence the size of $\mathsf{P}_a^{\microtree}$ is equal to the number of nodes
in $\microtree$ that have hard W-links for character $a$.
For all such characters $a$,
we store $\mathsf{P}_a^{\microtree}$ in $\microtree$.
The total size of these arrays for all the micro trees is clearly $O(n)$.

\begin{figure}[ph]
        \centering
         \includegraphics[scale=0.5]{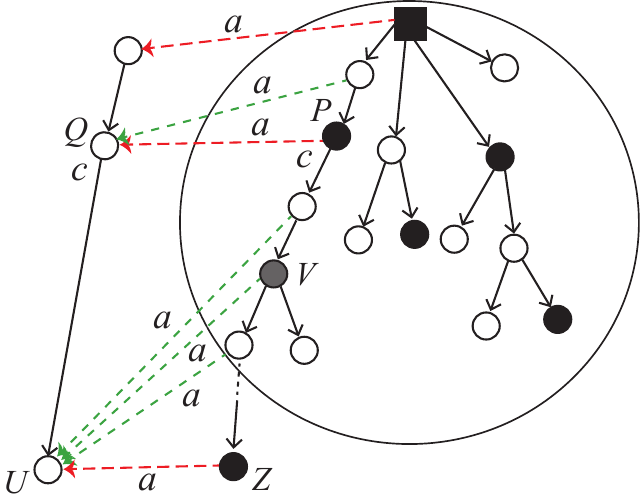}
	 \caption{Case (A)-(i)-(1) of our soft W-link query algorithm.}
         \label{fig:case-A-i-1}
\end{figure}

\begin{figure}[ph]
        \centering
         \includegraphics[scale=0.5]{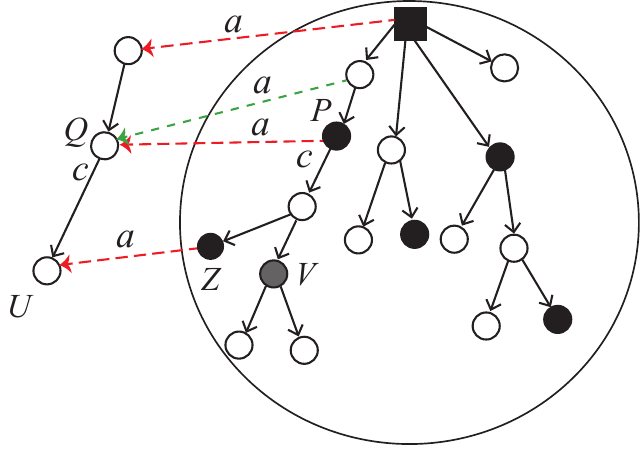}
	 \caption{Case (A)-(i)-(2) of our soft W-link query algorithm.}
         \label{fig:case-A-i-2}
\end{figure}

\begin{figure}[ph]
         \centering
         \includegraphics[scale=0.5]{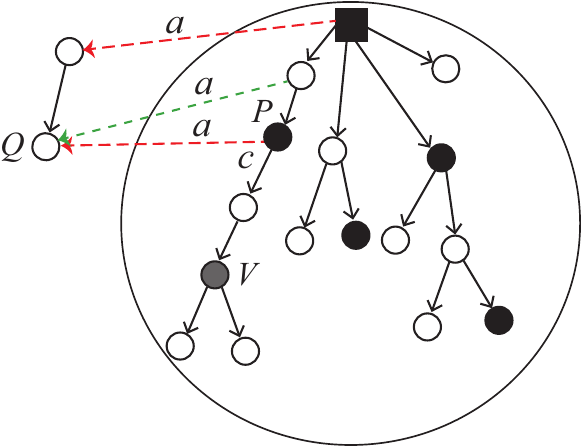}
	 \caption{Case (A)-(ii) of our soft W-link query algorithm.}
         \label{fig:case-A-ii}
\end{figure}

\begin{figure}[t!]
          \centering
         \includegraphics[scale=0.5]{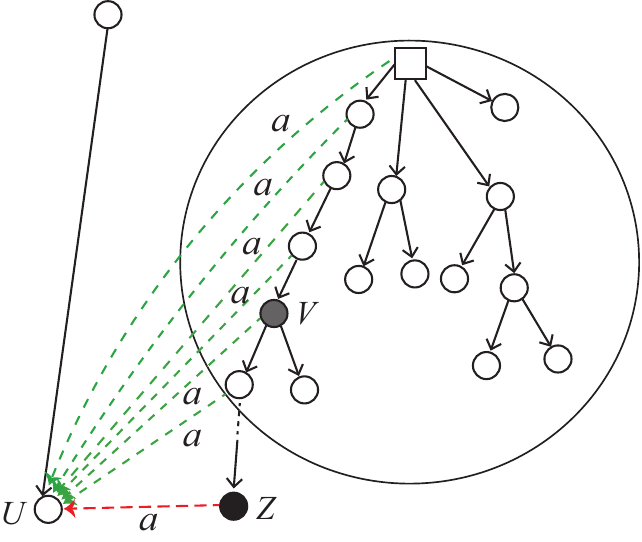}
	 \caption{Case (B)-(i) of our soft W-link query algorithm.}
         \label{fig:case-B-i}
\end{figure}

\begin{figure}[t!]
  \centering
         \includegraphics[scale=0.5]{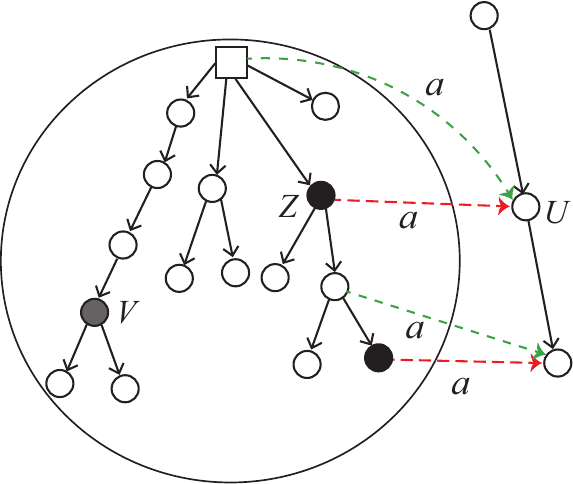}
	 \caption{Case (B)-(ii) of our soft W-link query algorithm.}
         \label{fig:case-B-ii} 
\end{figure}

Let $a \in \Sigma$ be any character,
and $V$ any node in $\STree(\BT)$ which does not have a hard W-link for $a$.
We wish to know if $V$ has a soft W-link for $a$,
and if so, we want to retrieve the target node of this link.
Let $\microtree$ denote the micro-tree to which $V$ belongs.
Consider the case where $V$ is not the root $R$ of $\microtree$,
since otherwise $\W{a}{V}$ is explicitly stored.
If $\W{a}{R}$ is nil, then by Fact~\ref{fact:soft_wlink_monotonicity}-(a) 
no nodes in the micro tree have W-links for character $a$.
Otherwise (if $\W{a}{R}$ exists), then we can find $\W{a}{V}$ as follows:
\begin{enumerate}
  \item[(A)] If there is an ancestor of $V$ that has a hard W-link with $a$ in the micro tree $\microtree$,
    then let $P$ be the lowest such ancestor of $V$.
    We follow the hard W-link $\W{a}{P}$ from $P$.
    Let $Q = \W{a}{P}$,
    and let $c$ be the first character in the path from $P$ to $V$.
    \begin{enumerate}
    \item[(i)]
      If $Q$ has an out-going edge whose label begins with $c$,
      then let $U$ be the child of $Q$ below this edge.
      \begin{enumerate}
      \item[(1)]
        If $Z = \suflink(U)$ is a descendant of $V$, then $U$ is the destination of
        the soft W-link $\W{a}{V}$ from $V$ for $a$ (see Figure~\ref{fig:case-A-i-1}).
      \item[(2)]
        Otherwise, there is no W-link from $V$ for $a$ (see Figure~\ref{fig:case-A-i-2}).
      \end{enumerate}
    \item[(ii)]
      Otherwise, there is no W-link from $V$ for $a$
      (see Figure~\ref{fig:case-A-ii}).
    \end{enumerate}  
  \item[(B)] Otherwise, $\W{a}{R}$ from the root $R$ of $\microtree$ is a soft W-link,
    which is explicitly stored. We follow it and let $U = \W{a}{R}$.
    \begin{enumerate}
    \item[(i)] If $Z = \suflink(U)$ is a descendant of $V$,
      then $U$ is the destination of the soft W-link $\W{a}{V}$ from $V$ for $a$
      (see Figure~\ref{fig:case-B-i}).
    \item[(ii)] Otherwise,
      there is no W-link from $V$ for $a$
      (see Figure~\ref{fig:case-B-ii}).
    \end{enumerate}
\end{enumerate}

In Figures~\ref{fig:case-A-i-1}, \ref{fig:case-A-i-2}, \ref{fig:case-A-ii}, \ref{fig:case-B-i} and~\ref{fig:case-B-ii},
the large circles show micro tree $\microtree$
and the square nodes are the roots of $\microtree$.
We query the soft W-link of $V$ (gray nodes) for character $a$. 
The black nodes are the nodes that have hard W-link for character $a$,
and the red broken arrows represent hard W-links for $a$ of our interest.
The green broken arrows represent soft W-links for $a$ of our interest.

Figures~\ref{fig:case-A-i-1}, \ref{fig:case-A-i-2}, and \ref{fig:case-A-ii} respectively show
the sub-cases of Case (A)-(i)-(1), Case (A)-(ii)-(2), and Case (A)-(ii)
where the root of the micro tree $\microtree$ has a hard W-link for $a$,
but our algorithm works also in the sub-cases where the root
has a soft W-link for $a$.

We remark that in Case (B) there can be
at most one path in the micro tree $\microtree$
containing nodes which have hard W-links for character $a$,
as illustrated in Figure~\ref{fig:case-B-i} and in Figure~\ref{fig:case-B-ii}.
This is because, if there are two distinct such paths in $\microtree$,
then by Fact~\ref{fact:soft_wlink_monotonicity}-(b)
the root of $\microtree$ must have a hard W-link for character $a$,
which contradicts our assumption for Case (B).

\begin{lemma} \label{lem:edge_simulation_DAWG}
  Our data structure simulates secondary-edge traversal on $\DAWG(\FT)$
  for a forward trie with $n$ nodes in $O(\log \sigma)$ time each,
  and occupies $O(n)$ space.
\end{lemma}

\begin{proof}
The correctness of this algorithm follows from
Fact~\ref{fact:soft_wlink_monotonicity}-(e).

%

Let us consider the time complexity.
The key is how to find the lowest ancestor $P$ for a given node $V$ and for a given character $a$ in the micro tree $\microtree$ to which $V$ belongs.

We show that at most two predecessor queries and one membership query
on $\mathsf{P}_a^{\microtree}$ are suffice to find $P$.
Let $P_1$ be the predecessor of $V$ in $\mathsf{P}_a^{\microtree}$ (if it exists),
and let $L_1 = \lca(V, P_1)$.
Let $P_2$ be the predecessor of $L_1$ in $\mathsf{P}_a^{\microtree}$ (if it exists),
and let $L_2 = \lca(V, P_2)$.

\begin{enumerate}
  \item If $P_1$ does not exist, then this case falls into Case (B). 
  \item If $P_1$ exists and it is an ancestor of $V$, then $P = P_1$ and this case falls into Case (A).
  \item Otherwise ($P_1$ exists but it is not an ancestor of $V$). Note that $L_1$ is in this micro-tree $\microtree$,
    since otherwise $P_1$ cannot be in the same micro-tree $\microtree$ as $V$.
    \begin{enumerate}
      \item If $L_1 \in \mathsf{P}_a^{\microtree}$, then $P = L_1$ and this case falls into Case (A).
      \item If $L_1 \notin \mathsf{P}_a^{\microtree}$:
        \begin{enumerate}
          \item If $P_2$ does not exist, then this case falls into Case (B). 
          \item If $P_2$ exists and it is an ancestor of $V$, then $P = P_2$ and this case falls into Case (A).
          \item Otherwise ($P_2$ exists but it is not an ancestor of $V$), then let $L_2 = \lca(P_1, P_2)$.
            Then, by Fact~\ref{fact:soft_wlink_monotonicity}-(b), $L_2$ has a hard W-link for $a$.
            Since $L_2$ is in this micro-tree $\microtree$, $P = L_2$ and this case falls into Case (A).
        \end{enumerate}
    \end{enumerate}
\end{enumerate}
See also Figure~\ref{fig:query_algorithm}
that illustrates Case 3-(b)-iii.

\begin{figure}[tb]
 \centering
 \includegraphics[scale=0.7]{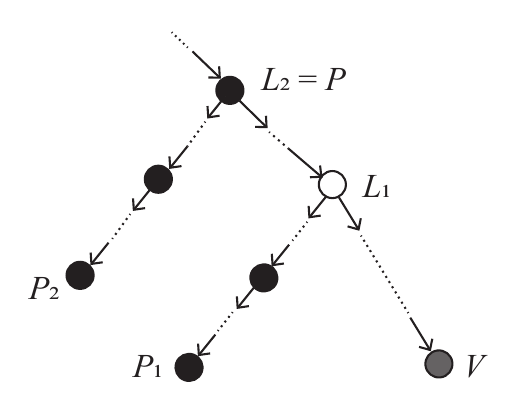}
 \caption{Illustration for our algorithm to find the lowest ancestor $P$ of a given node $V$ that has a hard W-link for character $a$. This figure shows Case 3-(b)-iii. The black nodes have hard W-links for $a$.}
 \label{fig:query_algorithm}
\end{figure}

We use the well-known LCA data structure~\cite{Bender}
that computes the LCA of two given nodes in $O(1)$ time after $O(n)$-time preprocessing.
Since each micro-tree contains $O(\sigma)$ nodes,
the size of $\mathsf{P}_a^{\microtree}$ is $O(\sigma)$ and thus 
predecessor queries and membership queries can be answered in $O(\log \sigma)$ time each by binary search.
We can check if one node is an ancestor of the other node
in $O(1)$ time, after standard $O(n)$-time preprocessing (or alternatively we can use LCA queries).
Consequently, this algorithm simulates soft W-links,
which are the secondary edges of the DAWG, in $O(\log \sigma)$ time.

We have already shown that the space complexity is $O(n)$.
\end{proof}

Note that our data structure for secondary-edge traversal on $\DAWG(\FT)$
is as time and space efficient as a standard representation
of labeled graphs where branches are implemented by balanced binary search trees
or arrays.

\subsection{Construction Algorithm}

What remains is how to preprocess the input trie to compute the above data structure.

\begin{lemma} \label{lem:computing_hard_Wlinks}
  Given a backward trie $\BT$ with $n$ nodes,
  we can compute $\STree(\BT)$ with all hard W-links in $O(n)$ time and space.
\end{lemma}

\begin{proof}
  We build $\STree(\BT)$ without suffix links
  in $O(n)$ time and space~\cite{Shibuya03}.
  We also perform a standard $O(n)$-time traversal on $\STree(\BT)$ so that
  we can answer in $O(1)$ time whether a node in $\STree(\BT)$ is an ancestor
  of another node in $\STree(\BT)$.

  We then add the suffix links to $\STree(\BT)$ as follows.
  To each node $v$ of $\BT$ we allocate its rank in a breadth-first traversal
  so that for any reversed edge $\Bpathstr(v, a, u)$,
  $v$ has a smaller rank than $u$.
  We will identify each node with its rank.
  
  Let $\SA$ be the suffix array for $\BT$ that corresponds to the leaves of $\STree(\BT)$,
  where $\SA[i] = j$ iff the suffix in $\BT$ beginning at node $j$
  is the $i$th lexicographically smallest suffix.
  We compute $\SA$ and its inverse array in $O(n)$ time via $\STree(\BT)$,
  or directly from $\BT$ using the algorithm proposed by Ferragina et al.~\cite{FerraginaLMM09}.
  The suffix links of the leaves of $\STree(\BT)$ can easily be computed
  in $O(n)$ time and space, by using the inverse array of $\SA$.
  Unlike the case of a single string where
  the suffix links of the leaves form a single chain,
  the suffix links of the leaves of $\STree(\BT)$ form a tree,
  but this does not incur any problem in our algorithm.
  To compute the suffix links of the internal nodes of $\STree(\BT)$,
  we use the following standard technique
  that was originally designed for the suffix tree of a single string
  (see e.g.~\cite{MBCT2015}):
  For any internal node $V$ in $\STree(\BT)$,
  let $\ell_V$ and $r_V$ denote the smallest and largest indices in $\SA$
  such that $\SA[\ell_V..r_V]$ is the maximal interval
  corresponding to the suffixes which have string $V$ as a prefix.
  It then holds that $\suflink(V) = U$,
  where $U$ is the lowest common ancestor (LCA) of $\suflink(\ell_V)$ and $\suflink(r_V)$.
  For all nodes $V$ in $\BT$,
  the LCA of $\suflink(\ell_V)$ and $\suflink(r_V)$ can be computed in $O(n)$ time and space.
  After computing the suffix links, we can easily compute the character labels
  of the corresponding hard W-links in $O(n)$ time.
\end{proof}

\begin{lemma} \label{lem:computing_soft_Wlinks}
  We can compute, in $O(n)$ time and space,
  all W-links of the macro tree nodes
  and the arrays $\mathsf{P}_a^{\microtree}$ for all the micro trees $\microtree$
  and characters $a \in \Sigma$.
\end{lemma}

\begin{proof}
  We perform a pre-order traversal on each micro tree $\microtree$.
  At each node $V$ visited during the traversal,
  we append the pre-order rank of $V$ to array $\mathsf{P}_a^{\microtree}$
  iff $V$ has a hard W-link $\W{a}{V}$ for character $a$.
  Since the size of $\microtree$ is $O(\sigma)$
  and since we have assumed an integer alphabet $[1..\sigma]$,
  we can compute $\mathsf{P}_a^{\microtree}$ for all characters $a$
  in $O(\sigma)$ time.
  It takes $O(\frac{n}{\sigma} \cdot \sigma) = O(n)$ time for all micro trees.

  The preprocessing for the macro tree consists of two steps.
  Firstly, 
  we need to compute soft W-links from
  the macro tree nodes
  (recall that we have already computed
  hard W-links from the macro tree nodes by Lemma~\ref{lem:computing_hard_Wlinks}).
  For this purpose, 
  in the above preprocessing for micro trees,
  we additionally pre-compute the successor of the root $R$ of
  each micro tree $\microtree$
  in each non-empty array $\mathsf{P}_a^{\microtree}$.
  By Fact~\ref{fact:soft_wlink_monotonicity}-(d),
  this successor corresponds to the unique descendant of $R$
  that has a hard W-link for character $a$.
  As above, this preprocessing also takes $O(\sigma)$ time for each micro tree,
  resulting in $O(n)$ total time.
%
  Secondly, we perform a bottom-up traversal on the macro tree.
  Our basic strategy is to ``propagate'' the soft W-links
  in a bottom up fashion from lower nodes to upper nodes in the macro tree
  (recall that these macro tree nodes are the roots of micro trees).
  In so doing, we first compute
  the soft W-links of the macro tree leaves.
  By Fact~\ref{fact:soft_wlink_monotonicity}-(c) and -(e),
  this can be done in $O(\sigma)$ time for each leaf using the successors computed above.
  Then we propagate the soft W-links to the macro tree internal nodes.
  The existence of soft W-links of internal nodes computed in this way
  is justified by Fact~\ref{fact:soft_wlink_monotonicity}-(a),
  however, the destinations of some soft W-links of some macro tree
  internal nodes may not be correct.
  This can happen when
  the corresponding micro trees contain hard W-links
  (due to Fact~\ref{fact:soft_wlink_monotonicity}-(e)).
  These destinations can be modified by using the successors of the roots
  computed in the first step,
  again due to Fact~\ref{fact:soft_wlink_monotonicity}-(e).
  Both of our propagation and modification steps take $O(\sigma)$ time
  for each macro tree node  of size $O(\sigma)$,
  and hence, it takes a total of $O(n)$ time.
\end{proof}

We have shown the following:
\begin{theorem} \label{theo:implicit_DAWG_FT}
  Given a forward trie $\FT$ of size $n$
  over an integer alphabet $\Sigma = [1..\sigma]$ with $\sigma = O(n)$,
  we can construct in $O(n)$ time and working space
  an $O(n)$-space representation of $\DAWG(\FT)$ that simulates
  edge traversals in $O(\log \sigma)$ time each.
\end{theorem}

\subsection{Bidirectional Searches on Tries}

A consequence of Theorem~\ref{theo:implicit_DAWG_FT}
is a space-efficient data structure that allows for \emph{bidirectional searches}
of patterns on a given trie.
\begin{theorem}
  Our $O(n)$-size implicit representation of
  Weiner links of $\STree(\BT)$ allows
  bidirectional searches of a given pattern $P$ of length $m$
  in $O(m \log \sigma + \occ)$ time,
  where $\occ$ is the number of occurrences of $P$
  in the input trie.
\end{theorem}

\begin{proof}
  We initially set $P \leftarrow \varepsilon$
  and start at the root of $\STree(\BT)$.

  Suppose we have added $k$~($0 \leq k < m$) characters to $P$
  in a bidirectional manner,
  and that we know the locus for the current $P$ over $\STree(\BT)$.
  
  If a new character $a$ is appended to $P$,
  then we perform a standard edge traversal on $\STree(\BT)$,
  in $O(\log \sigma)$ time.
  Then we set $P \leftarrow Pa$ and continue to the next character
  that will be either prepended or appended to $P$.

  If a new character $b$ is prepended to $P$,
  then there are two sub-cases.
  If the locus for $P$ is an explicit node of $\STree(\BT)$,
  then we perform the aforementioned algorithm
  and find the corresponding Weiner link for $b$,
  in $O(\log \sigma)$ time.
  Otherwise (if the locus for $P$ is an implicit node of $\STree(\BT)$),
  then let $V$ be the nearest explicit ancestor of the locus for $P$.
  We perform the aforementioned algorithm on $V$
  and find the destination $U$ of the Weiner link $\W{b}{V}$.
  Let $c$ be the first character of the edge from $V$ to the locus for $P$.
  We find the out-going edge of $U$ whose edge label begins with $c$
  in $O(\log \sigma)$ time.
  Then, the locus for $bP$ is at the $|P|-|V|$th character of this edge.
  Since each edge label in $\STree(\BT)$ is implemented
  by a pointer to a corresponding path in $\BT$,
  we can move to the locus for $bP$ in $O(1)$ time
  (using the level ancestor query on the trie).
  Then we set $P \leftarrow bP$ and continue to the next character
  that will be either prepended or appended to $P$.

  After adding all $m$ characters,
  we know the locus for the final pattern $P$ in $\STree(\BT)$.
  Then we use a standard technique of traversing
  the subtree rooted at the locus in $\STree(\BT)$, and takes all its leaves.
  Each of these leaves corresponds to the $\occ$ occurrences
  of $P$ in the trie.
  This takes $O(\occ)$ time.
\end{proof}

To the best of our knowledge, this is the \emph{first}
indexing structure on tries that permits efficient bidirectional pattern searches.

\section{Concluding Remarks}

This paper dealt with the labeled tree indexing problem.
We presented tight upper and lower bounds on
the sizes of the indexing structures
(i.e. suffix trees, DAWGs, CDAWGs, suffix arrays, affix trees, and affix arrays)
built on forward and backward tries.

In addition, we proposed a non-trivial $O(n)$-space implicit representation
of the DAWG for a forward trie for any alphabet of size $\sigma = O(n)$,
of which a na\"ive explicit representation must use $\Omega(n^2)$ space
in the worst case.
We also showed how to construct such a representation in $O(n)$ time
for an integer alphabet of size $\sigma = O(n)$.
The proposed data structure can also solve the bidirectional pattern search problem over a trie.

We believe that all these results have revealed a much clearer view
of the labeled tree indexing problem.
In addition, some of the quadratic lower bounds were generalized to
the labeled DAG indexing problem, which provided negative answers to the open questions
from the literature~\cite{MohriMW09,DenzumiTAM13}.

There remain intriguing open questions:
\begin{itemize}
   \item Direct construction of $\CDAWG(\BT)$: It is not difficult to construct $\CDAWG(\BT)$ from $\STree(\BT)$, but whether or not there is an efficient algorithm that builds $\CDAWG(\BT)$ \emph{directly} from a given backward trie $\BT$ remains open.

  \item Implicit representation of $\CDAWG(\FT)$: Can we generalize the ideas in Section~\ref{sec:DAWG_FT_construction} to $\CDAWG(\FT)$, so that it can be stored in $O(n)$ space for a large alphabet of size $\sigma = O(n)$?

  \item Bidirectional BWT for tries: Our bidirectional indexing structure for a trie is not succinct or compressed. Is there a version of BWT for tries that is capable of bidirectional searches with succinct or compressed space?

  \item Tight size bound on indexing structures for labeled DAGs: In Section~\ref{sec:labaled_DAG} we discussed that our quadratic lower bounds also hold for the DAWG or its SDD version built on a labeled DAG. We conjecture that a stronger lower bound holds for some labeled DAGs.
\end{itemize}

\section*{Acknowledgements}
The author thanks Dany Breslauer (deceased)
for fruitful discussions at the initial stage of this work.
The author is also grateful to an anonymous referee for pointing out an error in
the previous version of this paper and for suggesting to use micro-macro tree decomposition.
The author thanks Laurentius Leonard for his comments.

This research is supported by KAKENHI grant number JP17H01697
and by JST PRESTO Grant Number JPMJPR1922.

\bibliographystyle{abbrv}
\bibliography{ref}

\end{document}